\documentclass[final, journal, doublecolumn, 10pt, letterpaper]{IEEEtran}
\usepackage{amsmath,amssymb,amsfonts}
\usepackage{amsbsy}
\usepackage{graphicx}
\usepackage{graphics}
\usepackage{algorithm}
\usepackage[noend]{algorithmic}
\usepackage{subfigure, fmtcount}
\usepackage{cite}
\usepackage{multirow}
\usepackage{pdfsync}
\usepackage{epstopdf}
\usepackage{color}
\usepackage{enumerate}
\usepackage{paralist}
\usepackage{cite}
\usepackage{float}
\usepackage{wrapfig}
\usepackage{balance}
\usepackage{fancybox}
\usepackage[usenames,dvipsnames]{xcolor}
\usepackage{float}
\usepackage{wrapfig}
\usepackage{mathtools}

\makeatletter
 
\newcommand\floatc@simplerule[2]{{\@fs@cfont #1 #2}\par}
\newcommand\fs@simplerule{\def\@fs@cfont{\bfseries}\let\@fs@capt\floatc@simplerule
  \def\@fs@pre{\hrule height.8pt depth0pt \kern4pt}%
  \def\@fs@post{\kern4pt\hrule height.8pt depth0pt \kern4pt \relax}%
  \def\@fs@mid{\kern8pt}%
  \let\@fs@iftopcapt\iftrue}

\floatstyle{simplerule}
\newfloat{floatbox}{thp}{lob}
\floatname{floatbox}{Example}

\usepackage{amssymb}
\usepackage{amsmath}
\usepackage{amsfonts}
\usepackage{color}
\usepackage{balance}
\usepackage{paralist}
\usepackage{amsthm}
\newtheoremstyle{note}%
  {3pt}
  {0pt}
  {\itshape}
  {}
  {\bfseries}
  {:}
  {.5em}
  {}
\theoremstyle{note}

\newtheorem{example}{\textbf{Example}}

\newcommand{\transpose}{{\!\scriptscriptstyle\mathrm T}}
\newcommand{\norm}[1]{\lVert#1\rVert}
\newcommand{\Rbb}{\mathbb{R}}

\renewcommand{\l}{\ell}

\renewcommand{\L}{{\mathcal{L}}}
\newcommand{\G}{{\mathcal{G}}}
\newcommand{\E}{{\mathcal{E}}}

\newcommand{\V}{{\mathcal{V}}}
\newcommand{\D}{{\mathcal{D}}}
\newcommand{\Z}{{\mathcal{Z}}}
\newcommand{\N}{{\mathcal{N}}}

 \DeclareMathOperator*{\argmin}{argmin}

\long\def\comment#1{}

\newfont{\bbb}{msbm10 scaled 700}

\newfont{\bb}{msbm10 scaled 1100}

\newcommand{\RR}{\mbox{\bb R}}

\newcommand{\fv}{{\bf f}}

\newcommand{\Ec}{{\cal E}}

\newcommand{\Gc}{{\cal G}}

\newcommand{\Nc}{{\cal N}}
\newcommand{\Oc}{{\cal O}}

\newcommand{\Vc}{{\cal V}}

\newcommand{\Lcb}{\bf {\cal L}}

\newcommand{\psiv}{\hbox{\boldmath$\psi$}}

\definecolor{lightkhaki}{RGB}{250,250,210}
\definecolor{darkkhaki}{RGB}{139,129,76}
\begin{document}

\title{{\huge The Emerging Field of Signal Processing on Graphs}  \\ \emph{\Large Extending High-Dimensional Data Analysis to Networks and Other Irregular Domains} \vspace{.2in}}
\author{David I Shuman$^{\dag}$, Sunil K. Narang $^{\ddag}$, Pascal Frossard$^{\dag}$, Antonio Ortega$^{\ddag}$ and Pierre Vandergheynst$^{\dag}$\thanks{This work was supported in part by the European Commission under FET-Open grant number 255931 UNLocX and in part by the National Science Foundation (NSF) under grant CCF-1018977. The authors would like to thank the anonymous reviewers and Dorina Thanou for their constructive comments on earlier versions of this paper.}
 \\
\dag Ecole Polytechnique F\'ed\'erale de Lausanne (EPFL), Signal Processing Laboratory (LTS2 and LTS4) \\  
\ddag University of Southern California (USC), Signal and Image Processing Institute \\ 
\{david.shuman,~pascal.frossard,~pierre.vandergheynst\}@epfl.ch, kumarsun@usc.edu,~antonio.ortega@sipi.usc.edu}

\maketitle

\thispagestyle{empty}

\begin{abstract}
In applications such as social, energy, transportation, sensor, and neuronal networks, high-dimensional data naturally reside on the vertices of weighted graphs. The emerging field of signal processing on graphs merges algebraic and spectral graph theoretic concepts with computational harmonic analysis to process such signals on graphs. In this tutorial overview, we outline the main challenges 
of the area, discuss different ways to define graph spectral domains, which are the analogues to the classical frequency domain, and 
highlight the importance of incorporating the irregular structures of graph data domains when processing signals on graphs. 
We then review methods to generalize fundamental operations such as filtering, translation, modulation, dilation, and downsampling to the graph setting, and survey the localized, multiscale transforms that have been proposed to efficiently extract information from high-dimensional data on graphs. We conclude with a brief discussion of open issues and possible extensions.
\end{abstract}

\section{Introduction} \label{Se:intro}

\begin{figure}[h]
   \centerline{\includegraphics[width=.5\linewidth]{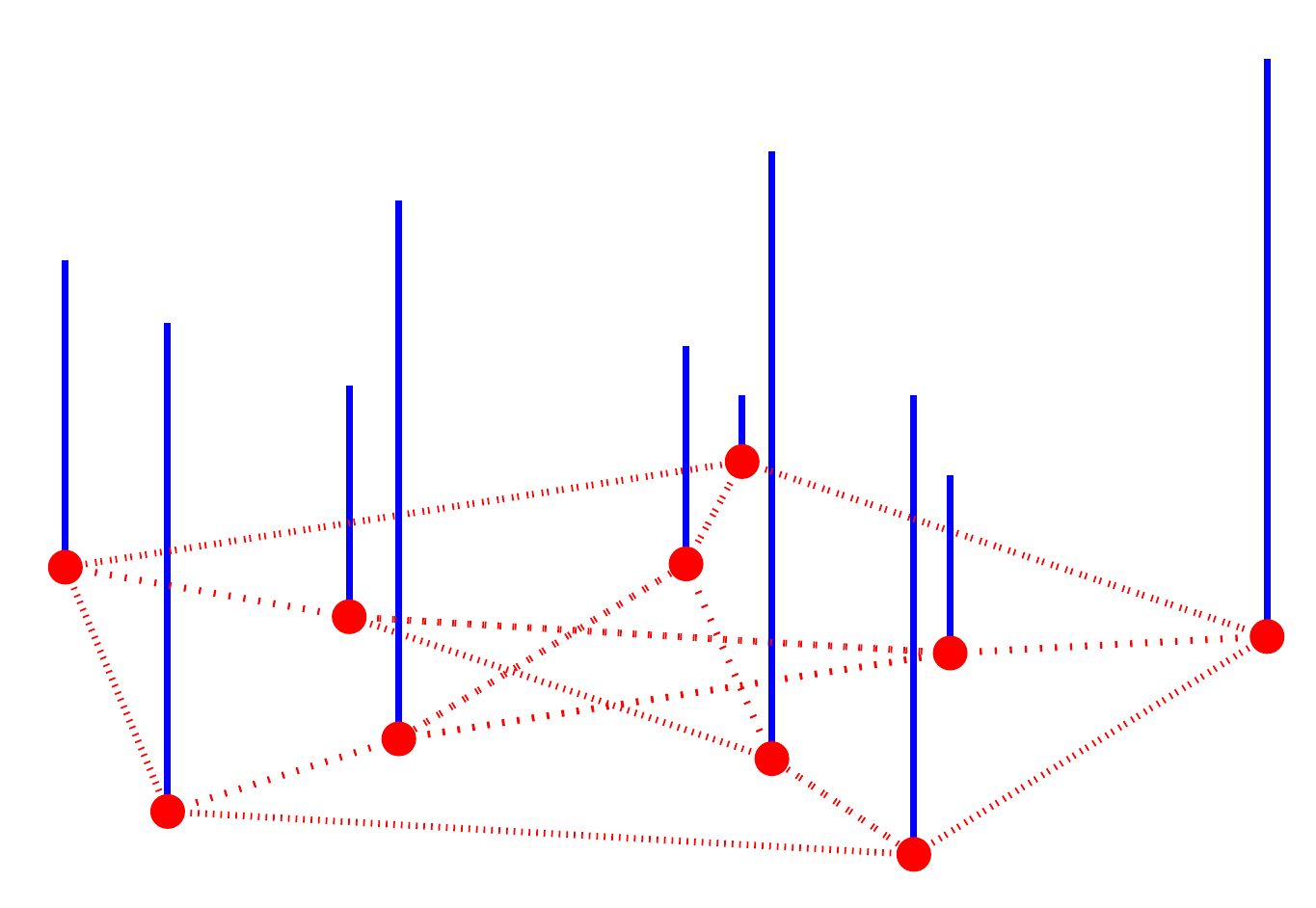}}
\caption {A random positive graph signal on the vertices of the Petersen graph. The height of each blue bar represents the signal value at the vertex
where the bar originates.}
  \label{Fig:petersen}
\end{figure}

Graphs are generic data representation forms which are useful for describing the geometric structures of data domains in numerous applications, including social, energy, transportation, sensor, and neuronal networks. 
The weight associated with each edge in the graph often represents the similarity between the two vertices it connects. The connectivities and edge weights are either dictated by the physics of the problem at hand or inferred from the data. For instance, the edge weight may be inversely proportional to the physical distance between nodes in the network. The data on these graphs can be visualized as a finite collection of samples, with one sample at each vertex in the graph. Collectively, we refer to these samples as a \emph{graph signal}. An example of a graph signal is shown in Figure \ref{Fig:petersen}. 

We find examples of graph signals in many different engineering and science fields.
In transportation networks, we may be interested in analyzing epidemiological data describing the spread of disease, census data describing human migration patterns, or logistics data describing inventories of trade goods (e.g. gasoline or grain stocks). 
In brain imaging, it is now possible to non-invasively infer the anatomical connectivity of distinct functional regions of the cerebral cortex \cite{hagmann}, and this connectivity can be represented by a weighted graph with the vertices corresponding to the functional regions of interest. Thus, noisy fMRI images can be viewed as signals on weighted graphs. 
Weighted graphs are also commonly used to represent similarities between data points in statistical learning problems for applications such as machine vision \cite{lowe} and automatic text classification \cite{apte}.
In fact, much of the literature on graph-based data analysis techniques emanates from the statistical learning community, as graph-based methods became especially popular for the semi-supervised learning problem where the objective is to classify unknown data with the help of a few labelled samples (e.g., \cite{smola,harmonic,zhou_scholkopf,reg_discrete,belkin_matveeva,zhou_bousquet}). In image processing, there has been a recent spike in graph-based filtering methods that build non-local and semi-local graphs to connect the pixels of the image based not only on their physical proximity, but also on noisy versions of the image to be processed  
(e.g., \cite{peyre_nlr,hancock,narang_SSP_2012} and references therein). Such methods are often able to better recognize and account for 
image edges and textures.

Common data processing tasks in these applications include
filtering, denoising, inpainting, and compressing graph signals. How can data be processed on irregular data domains such as arbitrary graphs? What are the best ways to efficiently extract information, either statistically or visually, from this high-dimensional data, for the purposes of storage, communication, and analysis? Is it possible to use operators or algorithms from the classical digital signal processing toolboxes? These are just a few of the questions that underlie the field of signal processing on graphs.

\subsection{The Main Challenges of Signal Processing on Graphs} \label{Se:challenges}

The ability of wavelet, time-frequency, curvelet and other localized transforms to sparsely represent different classes of 
high-dimensional data such as audio signals and images that lie on regular Euclidean spaces has led to a number of resounding successes in the aforementioned signal processing tasks (see, e.g., \cite[Section II]{rubinstein_dict_learning} for a recent survey of transform methods).

Both a signal on a graph with $N$ vertices and a classical discrete-time signal with $N$ samples can be viewed as vectors in $\Rbb^N$. However,  a major obstacle to the application of the classical signal processing techniques in the graph setting is that processing the graph signal in the same ways as a discrete-time signal ignores key dependencies arising from the irregular data domain.\footnote{Throughout, we refer to signal processing concepts for analog or discrete-time signals as ``classical,'' in order to differentiate them from concepts defined in the graph signal framework.}
Moreover, many extremely simple yet fundamental concepts that underlie classical signal processing techniques become significantly more challenging in the graph setting:

\begin{itemize}

\item To translate an analog signal $f(t)$ to the right by 3, we simply perform a change of variable and consider $f(t-3)$. 
However, it is not immediately clear what it means to translate  a graph signal
``to the right by 3.'' The change of variable technique will not work as there is no meaning to $f(\circ-3)$ in the graph setting. One naive option would be to simply label the vertices from 1 to $N$ and define $f(\circ-3):=f(\hbox{mod}(\circ-3,N))$, but
it is not particularly useful to define a generalized translation that depends heavily on the order in which we (arbitrarily) label the vertices.  
The unavoidable fact is that weighted graphs are irregular structures that lack a shift-invariant notion of translation.\footnote{The exception is the class of highly regular graphs such as a ring graph that have circulant graph Laplacians. Grady and Polimeni \cite[p.158]{grady} refer to such graphs as \emph{shift invariant} graphs.}

\item Modulating a signal on the real line by multiplying by a complex exponential corresponds to translation in the Fourier domain. 
However,  
the analogous spectrum in the graph setting is discrete and irregularly spaced, and it is therefore non-trivial to define an operator that corresponds to translation in the graph spectral domain. 

\item We intuitively downsample a discrete-time signal by deleting every other data point, for example. Yet, what does it mean to downsample the signal on the vertices of the graph shown in Figure \ref{Fig:petersen}?
There is not an obvious notion of ``every other vertex'' of a weighted graph. 

\item Even when we do fix a notion of downsampling, in order to create a multiresolution on graphs, we need a method to generate a coarser version of the graph
that somehow captures the structural properties embedded in the original graph.

\end{itemize}

In addition to dealing with the irregularity of the data domain, the graphs in the previously mentioned applications can feature a large number of vertices, and therefore many data samples. In order to scale well with the size of the data, signal processing techniques for graph signals should  employ localized operations that compute information about the data at each vertex by using data from a small neighborhood of vertices close to it in the graph.

Therefore, the overarching challenges of processing signals on graphs are
1) in cases where the graph is not directly dictated to us by the application, deciding how to construct a weighted graph that captures the geometric structure of the underlying data domain; 2) incorporating the graph structure into localized transform methods; 3) at the same time, leveraging invaluable intuitions developed from years of signal processing research on Euclidean domains; and 4) developing computationally efficient implementations of the localized transforms, in order to extract information from high-dimensional data on graphs and other irregular data domains. 

To address these challenges, the emerging field of signal processing on graphs merges 
algebraic and spectral graph theoretic concepts with computational harmonic analysis.
There is an extensive literature in both algebraic graph theory  
(e.g., \cite{godsil}) and spectral graph theory  
(e.g., \cite{chung,spielman_survey} and references therein); however, the bulk of the research prior to the past decade focused on analyzing the underlying graphs,
as opposed to signals on graphs. 

Finally, we should note that researchers have also designed localized signal processing techniques for other irregular data domains 
such as polygonal meshes and manifolds. 
This work includes, for example, 
low-pass filtering as a smoothing operation to enhance the 
overall shape of an object \cite{taubin}, 
transform coding based on spectral decompositions for the compression 
of geometry data \cite{karni}, and multiresolution 
representations of large meshes by decomposing one 
surface into multiple levels with different details \cite{guskov}. There is no doubt that such work has inspired and will continue to inspire new signal processing techniques in the graph setting.

\subsection{Outline of the Paper}

The objective of this paper is to offer 
a tutorial overview of the analysis of data on graphs from a signal processing perspective.
In the next section, 
we discuss different ways to encode the graph structure and define graph spectral domains, which are the analogues to the classical frequency domain. Section \ref{Se:ops} surveys some generalized operators on signals on graphs, such as filtering, translation, modulation, and downsampling. These operators form the basis for a number of 
localized, multiscale transform methods, which we review in Section \ref{Se:transforms}. We conclude with a brief mention of some open issues and possible extensions in Section \ref{Se:discussion}. 

\section{The Graph Spectral Domains} \label{Se:gsd}

Spectral graph theory has historically focused on constructing, analyzing, and manipulating graphs, as opposed to signals on graphs. It has proved particularly useful for  
the construction of expander graphs \cite{hoory}, graph visualization \cite[Section 16.7]{spielman_survey}, spectral clustering \cite{spectral_clustering}, graph coloring \cite[Section 16.9]{spielman_survey}, and numerous other applications in chemistry, physics, and computer science (see, e.g., \cite{cvetkovic_survey} for a recent review).

In the area of signal processing on graphs, spectral graph theory has been leveraged as a tool to define frequency spectra 
and expansion bases for graph Fourier transforms. In this section, we review some basic definitions and notations from spectral graph theory, with a focus on how it enables us to extend many of the important mathematical ideas and intuitions from classical Fourier analysis to the graph setting.

\subsection{Weighted Graphs and Graph Signals}
We are interested in analyzing signals defined on an undirected, connected, weighted graph $\G = \{\V,\E,\mathbf{W}\}$, which consists of a finite set of vertices $\V$ with $|\V|=N$, a
set of edges $\E$, and a weighted adjacency matrix $\mathbf{W}$. If there is an edge $e=(i,j)$ connecting vertices $i$ and $j$, the entry $W_{i,j}$ represents the weight of the edge; otherwise, $W_{i,j}=0.$ If the graph $\G$ is not connected and has $M$ connected components ($M>1$), we can separate signals on $\G$ into $M$ pieces corresponding to the $M$ connected components, and independently process the separated signals on each of the subgraphs. 

When the edge weights are not naturally defined by an application, one common way to define the weight of an edge connecting vertices $i$ and $j$ is via a thresholded Gaussian kernel weighting function:
\begin{align}\label{Eq:gkw}
W_{i,j}=
\begin{cases}
\exp\left({-\frac{[dist(i,j)]^2}{2\theta^2}}\right) &\mbox{if } dist(i,j) \leq \kappa \\
0 &\mbox{otherwise}
\end{cases},
\end{align}
for some parameters $\theta$ and $\kappa$. In \eqref{Eq:gkw}, $dist(i,j)$ may represent a physical distance between vertices $i$ and $j$, or the Euclidean distance between two feature vectors describing $i$ and $j$, the latter of which is especially common in graph-based semi-supervised learning methods. 
A second common method is to connect each vertex to its $k$-nearest neighbors based on the physical or feature space distances.
For other graph construction methods, see, 
e.g., \cite[Chapter 4]{grady}.

A signal or function $f: \V \rightarrow \Rbb$ defined on the vertices of the graph may be represented as a vector $\mathbf{f} \in \Rbb^N$, where the $i^{th}$ component of the vector $\mathbf{f}$ represents the function value at the $i^{th}$ vertex in $\V$.\footnote{In order to analyze data residing on the \emph{edges} of an unweighted graph, one option is to build its \emph{line graph}, where we associate a vertex to each edge and connect two vertices in the line graph if their corresponding edges in the original graph share a common vertex, and then analyze the data on the vertices of the line graph.} The \emph{graph signal} in Figure \ref{Fig:petersen} 
is one such example.

\subsection{The Non-Normalized Graph Laplacian}
The \emph{non-normalized graph Laplacian}, also called the combinatorial graph Laplacian, is defined as $\L:=\mathbf{D}-\mathbf{W}$, where the degree matrix $\mathbf{D}$ is a diagonal matrix whose $i^{th}$ diagonal element $d_i$ is equal to the sum of the weights of all the edges incident to vertex $i$. The graph Laplacian is a difference operator, as, for any signal $\mathbf{f} \in \Rbb^N$, it satisfies
\begin{align*}
(\L f)(i)=\sum_{j \in \N_i} W_{i,j}[f(i)-f(j)],
\end{align*}
where the neighborhood $\N_i$ is the set of vertices connected to vertex $i$ by an edge. More generally, we denote by $\N(i,k)$ the set of vertices  
connected to vertex $i$ by a path of $k$ or fewer edges.

Because the graph Laplacian $\L$ is a real symmetric
matrix, it has a complete set of  
orthonormal eigenvectors, which we denote 
by $\left\{\mathbf{u}_{\l}\right\}_{\l=0,1,\ldots,N-1}$.\footnote{Note that there is not necessarily a unique set of graph Laplacian eigenvectors, but we assume throughout that a set of eigenvectors is chosen and fixed.} These eigenvectors have associated real, non-negative eigenvalues
$\left\{\lambda_{\l}\right\}_{\l=0,1,\ldots,N-1}$ satisfying $\L{\mathbf{u}}_{\l} = \lambda_{\l}{\mathbf{u}}_{\l}$, for $\l=0,1,\ldots,N-1$. Zero appears as an eigenvalue with multiplicity equal to the
number of connected components of the graph
\cite{chung}, and thus, since we consider connected graphs, we assume the graph Laplacian eigenvalues are ordered as
$0=\lambda_0 < \lambda_1 \leq \lambda_2 ... \leq \lambda_{N-1}:=\lambda_{\max}$. We denote the entire spectrum by $\sigma(\L):=\{\lambda_0,\lambda_1,\ldots,\lambda_{N-1}\}$.

\subsection{A Graph Fourier Transform and Notion of Frequency} \label{Se:frequency}
The classical Fourier transform 
\begin{align*}
\hat{f}(\xi):= \langle f, e^{2\pi i \xi t}\rangle = \int\limits_{\Rbb}f(t)  e^{-2 \pi i \xi t}dt
\end{align*}
is the expansion of a function $f$ in terms of the complex exponentials, which are the eigenfunctions of the one-dimensional Laplace operator:
\begin{align}\label{Eq:classical_Lap_eigen}
-\Delta(e^{2\pi i \xi t})=-\frac{\partial^2}{\partial t^2} e^{2\pi i \xi t} = (2\pi\xi)^2 e^{2\pi i \xi t}.
\end{align}
Analogously,
we can define the \emph{graph Fourier transform} $\hat{\mathbf{f}}$ of any function $\mathbf{f}\in\mathbb{R}^{N}$ on the vertices of $\G$
as the expansion of $\mathbf{f}$ in terms of the eigenvectors of the graph Laplacian:
\begin{align}\label{Eq:graph_FT}
\hat{f}(\lambda_{\l}) := \langle {\mathbf{f},\mathbf{u}_{\l}}\rangle = \sum_{i=1}^N f(i) u_{\l}^*(i). 
\end{align}
The \emph{inverse graph Fourier transform} is then given by
\begin{align}\label{Eq:graph_IFT}
f(i) = \sum_{\l=0}^{N-1} \hat{f}(\lambda_{\l}) u_{\l}(i).
\end{align}

\begin{figure}[h]
\hfill
\begin{minipage}[b]{.30\linewidth}
   \centering
   \centerline{\includegraphics[width=\linewidth]{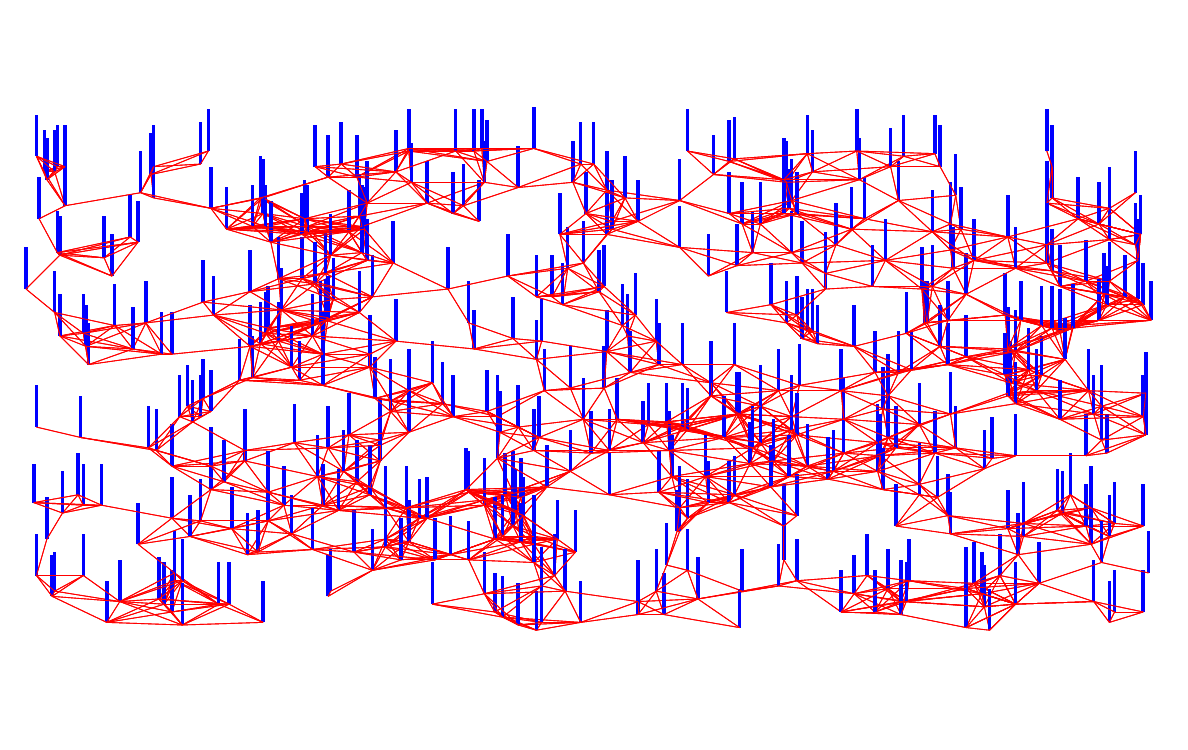}}
\centerline{\small{$\mathbf{u}_0$}}
\end{minipage}
\hfill
\begin{minipage}[b]{.30\linewidth}
   \centering
   \centerline{\includegraphics[width=\linewidth]{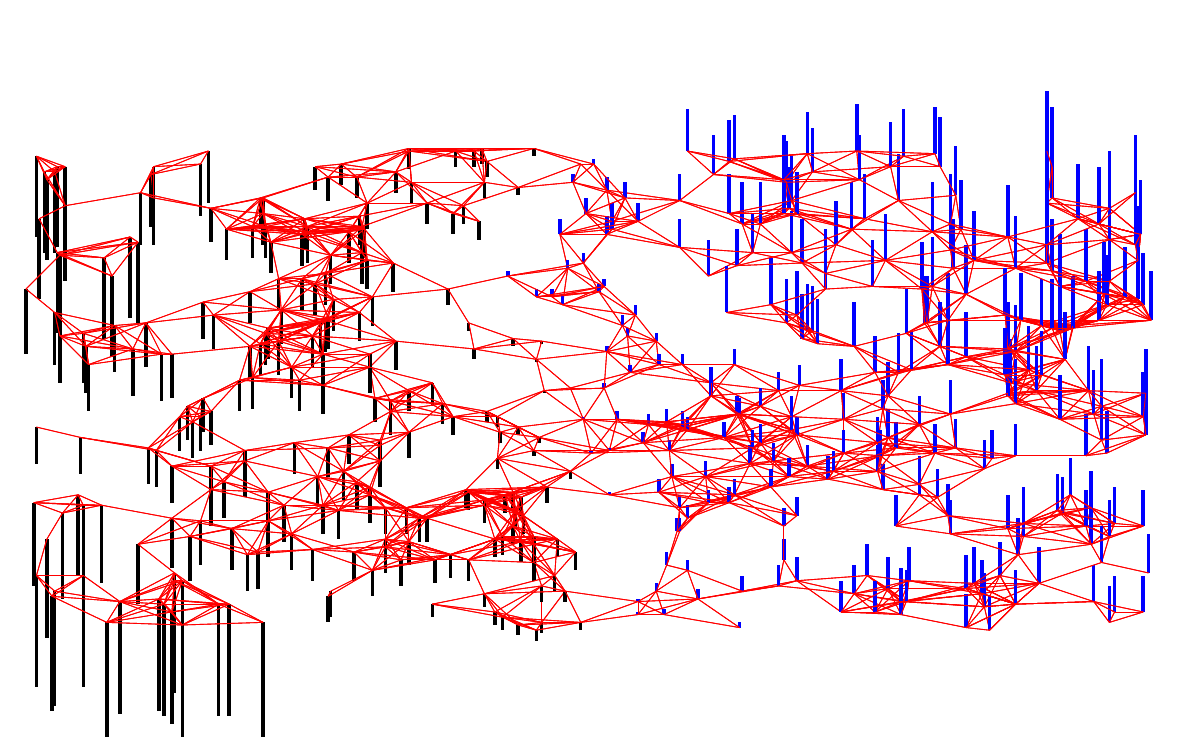}}
\centerline{\small{$\mathbf{u}_1$}}
\end{minipage}
\hfill
\hfill
\begin{minipage}[b]{.30\linewidth}
   \centering
   \centerline{\includegraphics[width=\linewidth]{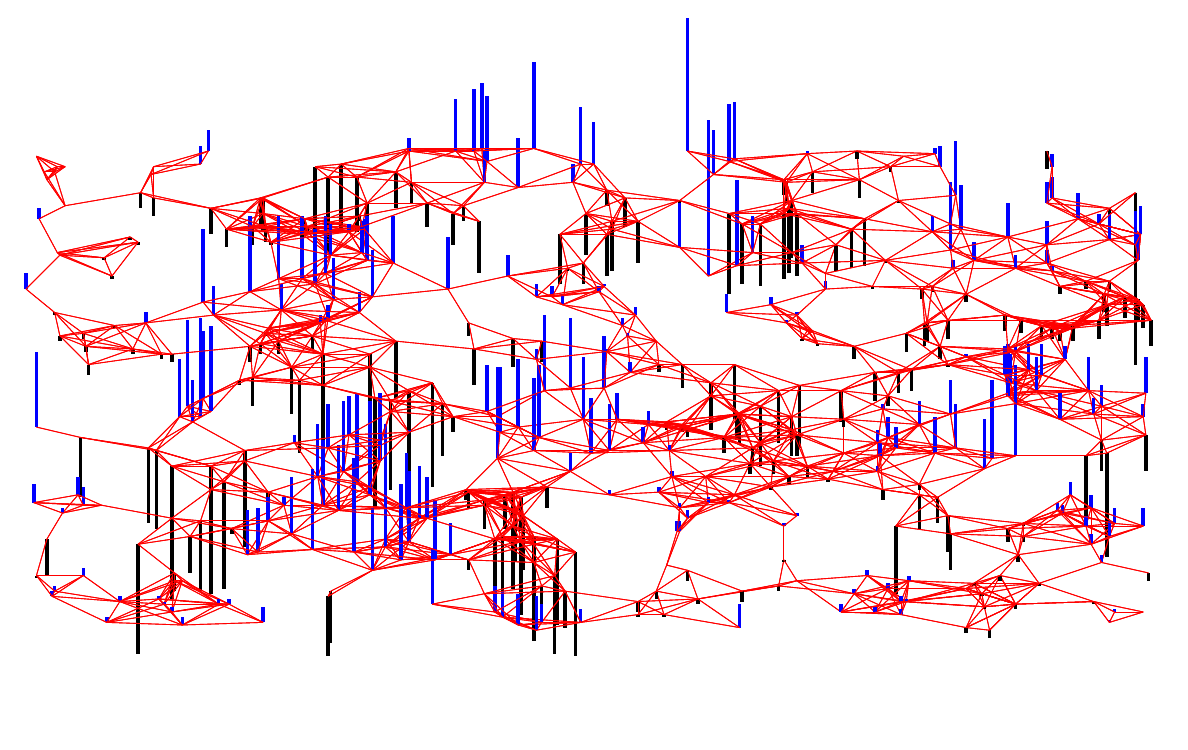}}
\centerline{\small{$\mathbf{u}_{50}$}}
\end{minipage}
\hfill
\hfill
\caption {Three graph Laplacian eigenvectors of a random sensor network graph. 
The signals' component values are represented by the blue (positive) and black (negative) bars coming out of the vertices. 
Note that $\mathbf{u}_{50}$ contains many more 
zero crossings than the constant eigenvector $\mathbf{u}_0$ and the smooth \emph{Fiedler vector} $\mathbf{u}_1$.}
\label{Fig:graph_lap_vecs}
\end{figure}

\begin{figure}[h]
\hfill
\begin{minipage}[b]{.42\linewidth}
   \centering
   \centerline{\includegraphics[width=\linewidth]{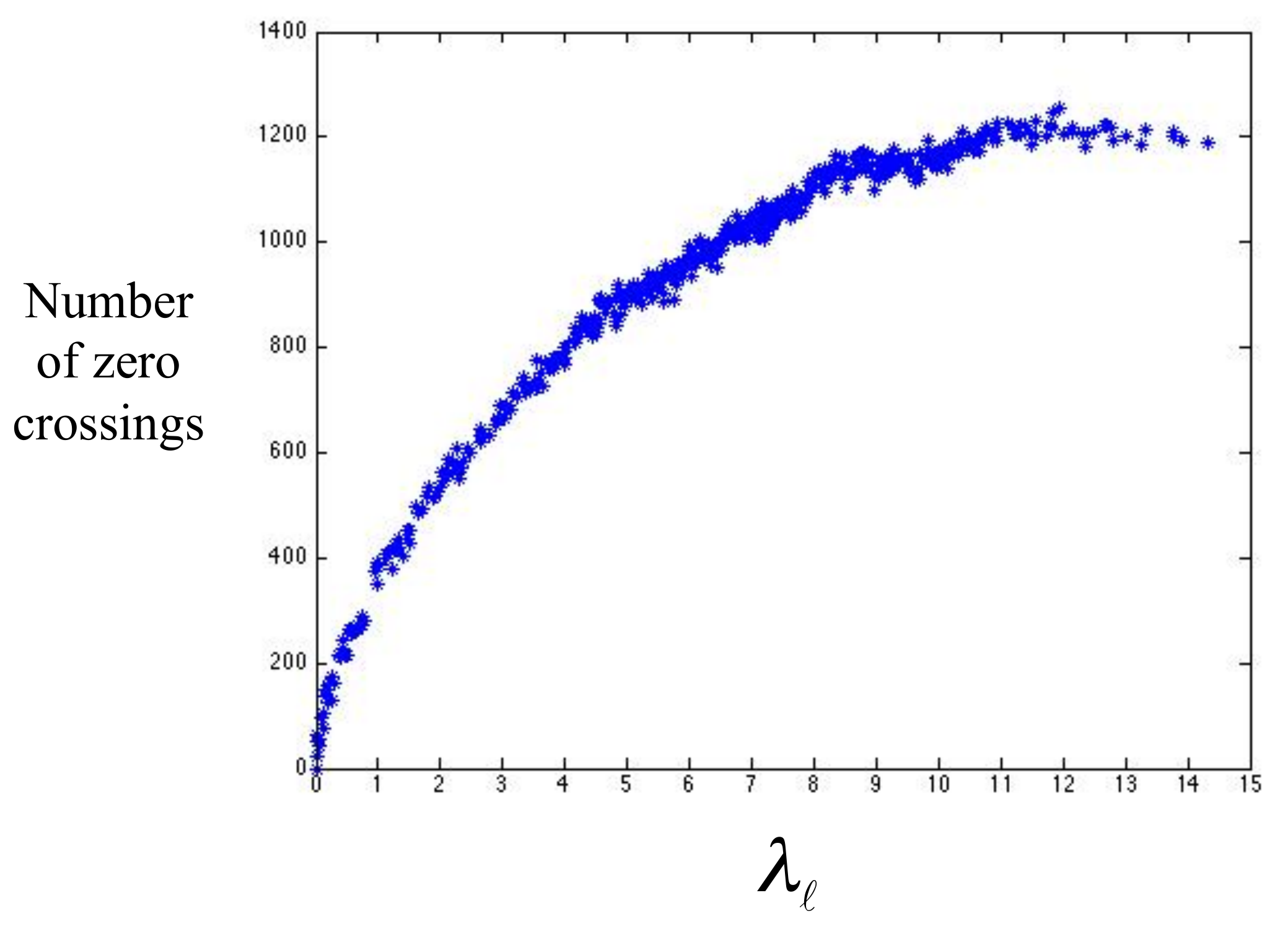}}
\centerline{\small{~~~~~~~(a)}}
\end{minipage}
\hfill
\begin{minipage}[b]{.42\linewidth}
   \centering
   \centerline{\includegraphics[width=\linewidth]{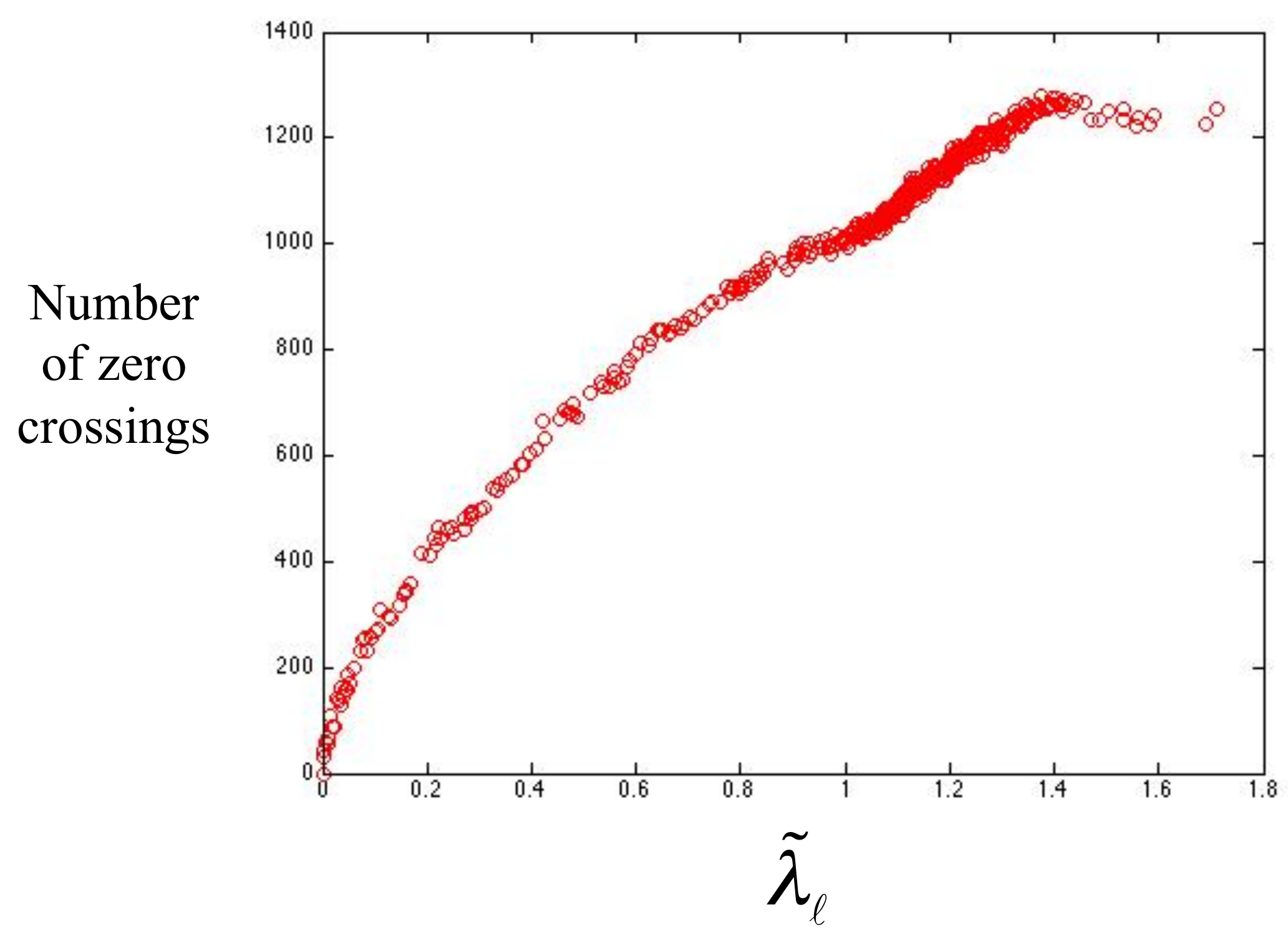}}
\centerline{\small{~~~~~~~~(b)}}
\end{minipage}
\hfill
\hfill
\caption {The number of zero crossings, $|\Z_{\G}(\mathbf{u}_{\l})|$ in (a) and $|\Z_{\G}(\tilde{\mathbf{u}}_{\l})|$ in (b), of the non-normalized and normalized graph Laplacian eigenvectors for the random sensor network graph of Figure \ref{Fig:graph_lap_vecs}, respectively (the latter of which is defined in Section \ref{Se:other_graph}). In both cases, the Laplacian eigenvectors associated with larger eigenvalues cross zero more often, confirming the interpretation of the graph Laplacian eigenvalues as notions of frequency.}
  \label{Fig:zero_crossings}
\end{figure}

In classical Fourier analysis, the eigenvalues $\{(2\pi\xi)^2\}_{\xi \in \Rbb}$ in \eqref{Eq:classical_Lap_eigen} carry a specific notion of frequency: for $\xi$ close to zero (low frequencies), the associated complex exponential eigenfunctions are smooth, slowly oscillating functions, whereas for $\xi$ far from zero (high frequencies), the associated complex exponential eigenfunctions oscillate much more rapidly.
In the graph setting, the graph Laplacian eigenvalues and eigenvectors provide a similar notion of frequency. For connected graphs, the Laplacian eigenvector $\mathbf{u}_0$ associated with the eigenvalue $0$ is constant and equal to $\frac{1}{\sqrt{N}}$ at each vertex. The graph Laplacian eigenvectors associated with low frequencies $\lambda_{\l}$ vary slowly across the graph; i.e., if two vertices are connected by an edge with a large weight, the values of the eigenvector at those locations are likely to be similar. The eigenvectors associated with larger eigenvalues oscillate more rapidly and are more likely to have dissimilar values on vertices connected by an edge with high weight. This is demonstrated in both Figure \ref{Fig:graph_lap_vecs}, which shows different graph Laplacian eigenvectors for a random sensor network graph, and Figure \ref{Fig:zero_crossings}, which shows the number $|\Z_{\G}(\cdot)|$ of \emph{zero crossings} of each graph Laplacian eigenvector. The set of zero crossings of a signal $\mathbf{f}$ on a graph $\G$ is defined as
\begin{align*}
\Z_{\G}(\mathbf{f}):=\left\{e=(i,j) \in \E : f(i)f(j)<0 \right\};
\end{align*}
that is, the set of edges connecting a vertex with a positive signal to a vertex with a negative signal.

\subsection{Graph Signal Representations in Two Domains}
\begin{figure}[h]
\hfill
\begin{minipage}[b]{.45\linewidth}
   \centering
   \centerline{\includegraphics[width=\linewidth]{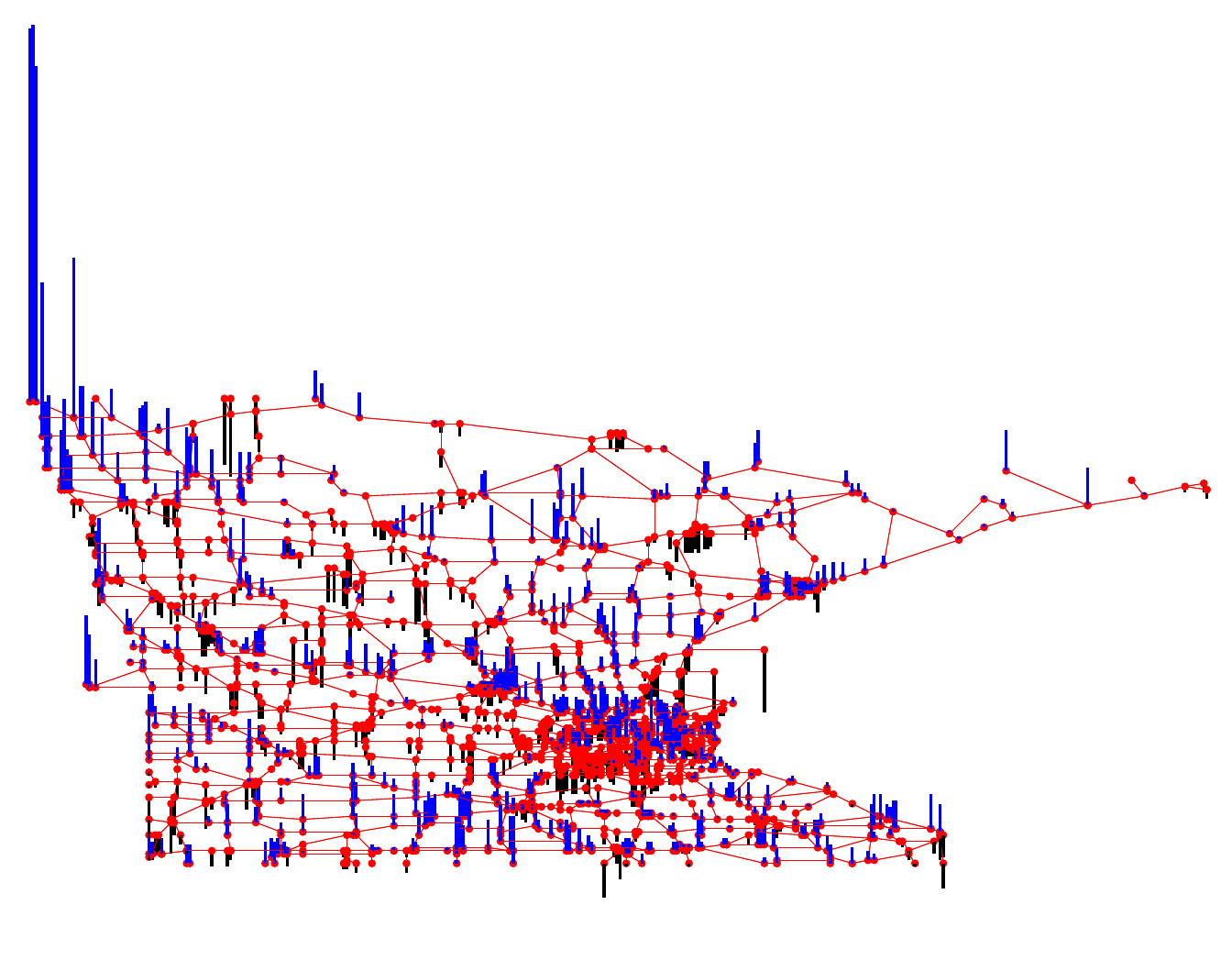}}
\centerline{\small{(a)~~~~~}}
\end{minipage}
\hfill
\begin{minipage}[b]{.45\linewidth}
   \centering
   \centerline{\includegraphics[width=\linewidth]{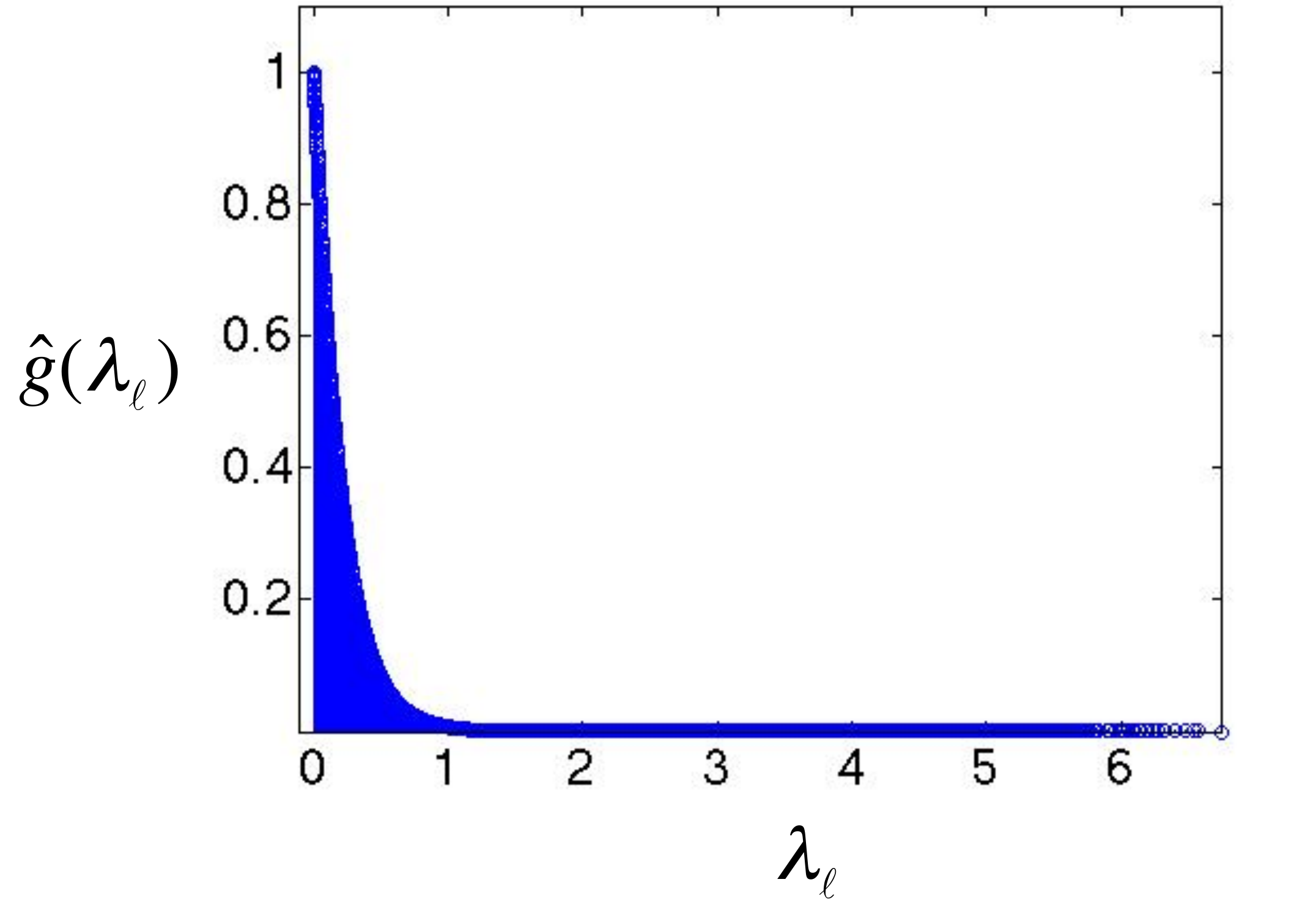}}
\centerline{\small{~~~~~~~~(b)}}
\end{minipage}
\hfill
\hfill
\caption{Equivalent representations of a graph signal in the vertex and graph spectral domains. (a) A signal $\mathbf{g}$ that resides on the vertices of the Minnesota road graph \cite{gleich} with Gaussian edge weights as in \eqref{Eq:gkw}. The signal's component values are represented by the blue (positive) and black (negative) bars coming out of the vertices. (b) The same signal in the graph spectral domain. In this case, the signal is a \emph{heat kernel} which is actually defined directly in the graph spectral domain by $\hat{g}(\lambda_{\l})=e^{-5\lambda_{\l}}$. The signal plotted in (a) is then determined by taking an inverse graph Fourier transform \eqref{Eq:graph_IFT} of $\hat{\mathbf{g}}$.}
 \label{Fig:minn_signals}
\end{figure}

The graph Fourier transform \eqref{Eq:graph_FT} and its inverse \eqref{Eq:graph_IFT} give us a way to equivalently represent a signal in two different domains: the vertex domain and the graph spectral domain. While we often start with a signal $\mathbf{g}$ in the vertex domain, it may also be useful to define a signal $\hat{\mathbf{g}}$ directly in the graph spectral domain. We refer to such signals as \emph{kernels}. In Figures \ref{Fig:minn_signals}(a) and \ref{Fig:minn_signals}(b), one such signal, a heat kernel, is shown in both domains. Analogously to the classical analog case, the graph Fourier coefficients of a smooth signal such as the one shown in Figure \ref{Fig:minn_signals} decay rapidly. Such signals are \emph{compressible} as they can be closely approximated by just a few graph Fourier coefficients (see, e.g., \cite{diffusion_wavelets, zhu_rabbat1,zhu_rabbat2} for ways to exploit this compressibility). 

\subsection{Discrete Calculus and Signal Smoothness with Respect to the Intrinsic Structure of the Graph}
When we 
analyze signals, it is important to emphasize that 
properties such as smoothness 
are \emph{with respect to the intrinsic structure of the data domain}, which in our context is the weighted graph. Whereas differential geometry provides  
tools to incorporate the geometric structure of the underlying manifold into the analysis of continuous signals on   
differentiable manifolds, 
\emph{discrete calculus} provides a ``set of definitions and differential operators that make it possible to operate the machinery of multivariate calculus on a finite, discrete space \cite[p. 1]{grady}.''  

To add mathematical precision to 
the notion of smoothness with respect to the intrinsic structure of the underlying graph, we briefly present some of the discrete
 differential operators defined in \cite{smola, zhou_scholkopf, reg_discrete, grady, belkin_matveeva, bougleux,elmoataz,osher_chapter}.\footnote{Note that the names of many of the discrete calculus 
operators correspond to the analogous operators in the continuous setting. In some problems, the weighted graph arises from a discrete sampling of a 
smooth manifold. In that situation, the discrete differential operators may converge -- possibly under additional assumptions -- to their namesake continuous operators as the density of the sampling increases. For example, \cite{belkin_conv} -\nocite{hein}\nocite{singer}\cite{ting} examine the convergence of discrete graph Laplacians (normalized and non-normalized) to continuous manifold Laplacians.} 
 The \emph{edge derivative} of a signal $\mathbf{f}$ with respect to edge $e=(i,j)$ at vertex $i$ is defined as
\begin{align*}
\left.\frac{\partial \mathbf{f}}{\partial e} \right|_i:=\sqrt{W_{i,j}}\left[f(j)-f(i)\right],
\end{align*}
and the \emph{graph gradient} of $\mathbf{f}$ at vertex $i$ is the vector
\begin{align*}
\triangledown_i \mathbf{f} := \left[\left\{\left.\frac{\partial \mathbf{f}}{\partial e} \right|_i\right\}_{e\in {\E}\hbox{ s.t. } e=(i,j) \hbox{ for some }j\in \V}\right].
\end{align*}
Then the \emph{local variation} at vertex $i$ 
\begin{align*}
\norm{\triangledown_i \mathbf{f}}_2 &:=\left[\sum_{e\in {\E}\hbox{ s.t. } e=(i,j) \hbox{ for some }j\in \V} \left(\left.\frac{\partial \mathbf{f}}{\partial e} \right|_i\right)^2\right]^{\frac{1}{2}} \\
&~=\left[\sum_{j \in \N_i}W_{i,j}\left[f(j)-f(i)\right]^2\right]^{\frac{1}{2}}
\end{align*}
provides a measure of local smoothness of $\mathbf{f}$ around vertex $i$, as it is small when the function $\mathbf{f}$ has similar values at $i$ and all
 neighboring vertices of $i$. 

For notions of global smoothness,
the \emph{discrete p-Dirichlet form} of $\mathbf{f}$ is defined as
\begin{align}\label{Eq:pdirichlet}
S_p(\mathbf{f}):=\frac{1}{p}\sum_{i \in V}\norm{\triangledown_i \mathbf{f}}_2^p=\frac{1}{p}\sum_{i \in V}\left[\sum_{j \in \N_i}W_{i,j}\left[f(j)-f(i)\right]^2\right]^{\frac{p}{2}}.
\end{align}
When $p=1$, $S_1(\mathbf{f})$ is the \emph{total variation} of the signal with respect to the graph. When $p=2$, we have
\begin{align}\label{Eq:s2def}
S_2(\mathbf{f})&=\frac{1}{2}\sum_{i \in V}\sum_{j \in \N_i}W_{i,j}\left[f(j)-f(i)\right]^2 \nonumber \\
&=\sum_{(i,j)\in {\E}}W_{i,j}\left[f(j)-f(i)\right]^2=\mathbf{f}^{\transpose}\L \mathbf{f}.
\end{align}
$S_2(\mathbf{f})$ is known as the \emph{graph Laplacian quadratic form} \cite{spielman_survey}, and the semi-norm
$\norm{\mathbf{f}}_{\L}$ is defined as 
\begin{align*}
\norm{\mathbf{f}}_{\L}:=\norm{\L^{\frac{1}{2}} \mathbf{f}}_2=\sqrt{\mathbf{f}^{\transpose}\L \mathbf{f}}=\sqrt{S_2(\mathbf{f})}.
\end{align*}
Note from \eqref{Eq:s2def} that the quadratic form $S_2(\mathbf{f})$ is equal to zero if and only if $\mathbf{f}$ is constant across all vertices (which is why $\norm{\mathbf{f}}_{\L}$ is only a semi-norm), and, more generally, $S_2(\mathbf{f})$ is small when the signal $\mathbf{f}$ has similar values at neighboring vertices connected by an edge with a large weight; i.e., when it is smooth. 

Returning to the graph Laplacian eigenvalues and eigenvectors, the Courant-Fischer Theorem \cite[Theorem 4.2.11]{horn} tells us they can also be defined iteratively via the Rayleigh quotient as
\begin{align}\label{Eq:rayleigh1}
\lambda_0&=\min_{
\substack{\mathbf{f}\in \Rbb^N \\ \norm{\mathbf{f}}_2=1}
} 
\left\{{\mathbf{f}^{\transpose}\L \mathbf{f}}\right\}, 
\\
\hbox{ and } 
\lambda_{\l}&=~\smashoperator{\min_{
\substack{ 
\mathbf{f}\in \Rbb^N \\ 
\norm{\mathbf{f}}_2=1 \\ 
\mathbf{f} \perp span\{\mathbf{u}_{0},\ldots,\mathbf{u}_{\l-1} \}
}}
}
~\left\{{\mathbf{f}^{\transpose}\L \mathbf{f}} \right\},~\l=1,2,\ldots,N-1, \label{Eq:rayleigh2}
\end{align} 
where the eigenvector $\mathbf{u}_{\l}$ is the minimizer of the $\l^{th}$ problem. 
From \eqref{Eq:s2def} and \eqref{Eq:rayleigh1}, we see again why 
$\mathbf{u}_0$ is constant for connected graphs.
Equation \eqref{Eq:rayleigh2} explains
why the graph Laplacian eigenvectors associated with lower eigenvalues are smoother, and provides another interpretation for why the graph Laplacian spectrum carries a notion of frequency.  

\begin{figure}[h] 
\fboxsep=3mm
\fboxrule=2pt
\fcolorbox{darkkhaki}{lightkhaki}{
\begin{minipage}{3.08in}
\hfill
\begin{minipage}[b]{.30\linewidth}
   \centering
   \centerline{\small{~~~~$\G_1$}}
   \centerline{\includegraphics[width=\linewidth]{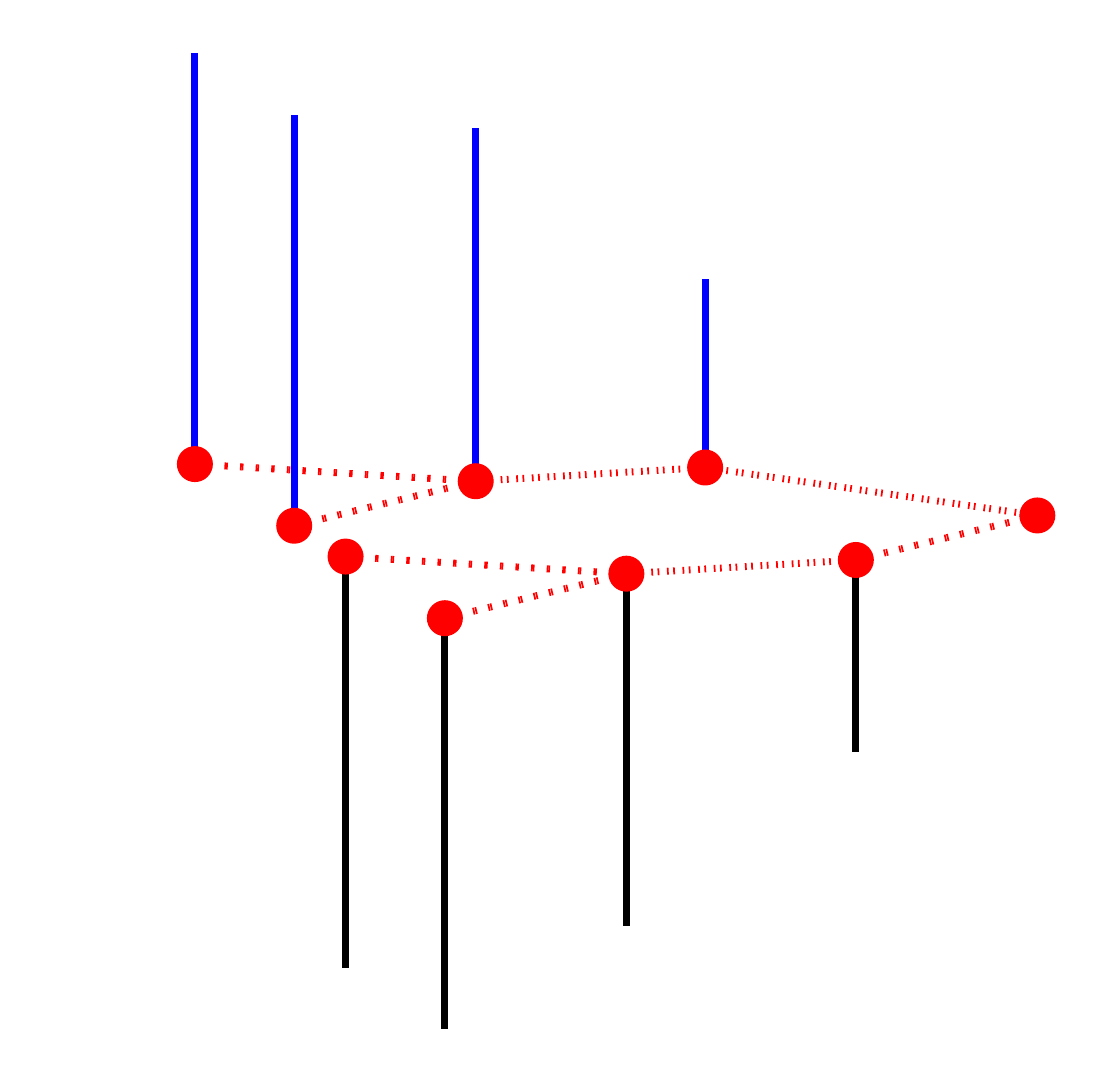}} 
   \centerline{\includegraphics[width=1.05\linewidth]{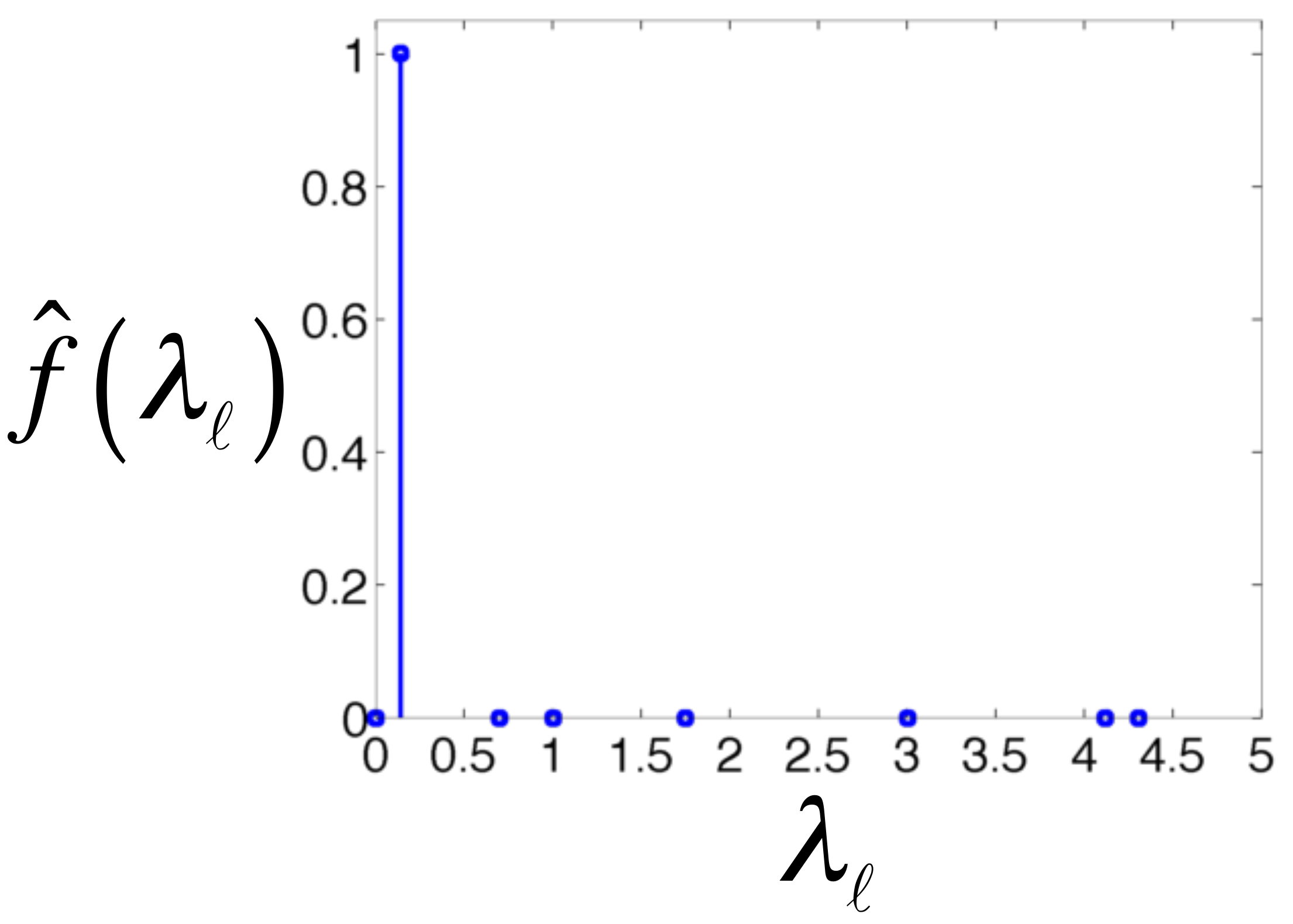}}

\end{minipage}
\hfill
\begin{minipage}[b]{.30\linewidth}
   \centering
   \centerline{\small{~~~~$\G_2$}}
   \centerline{\includegraphics[width=\linewidth]{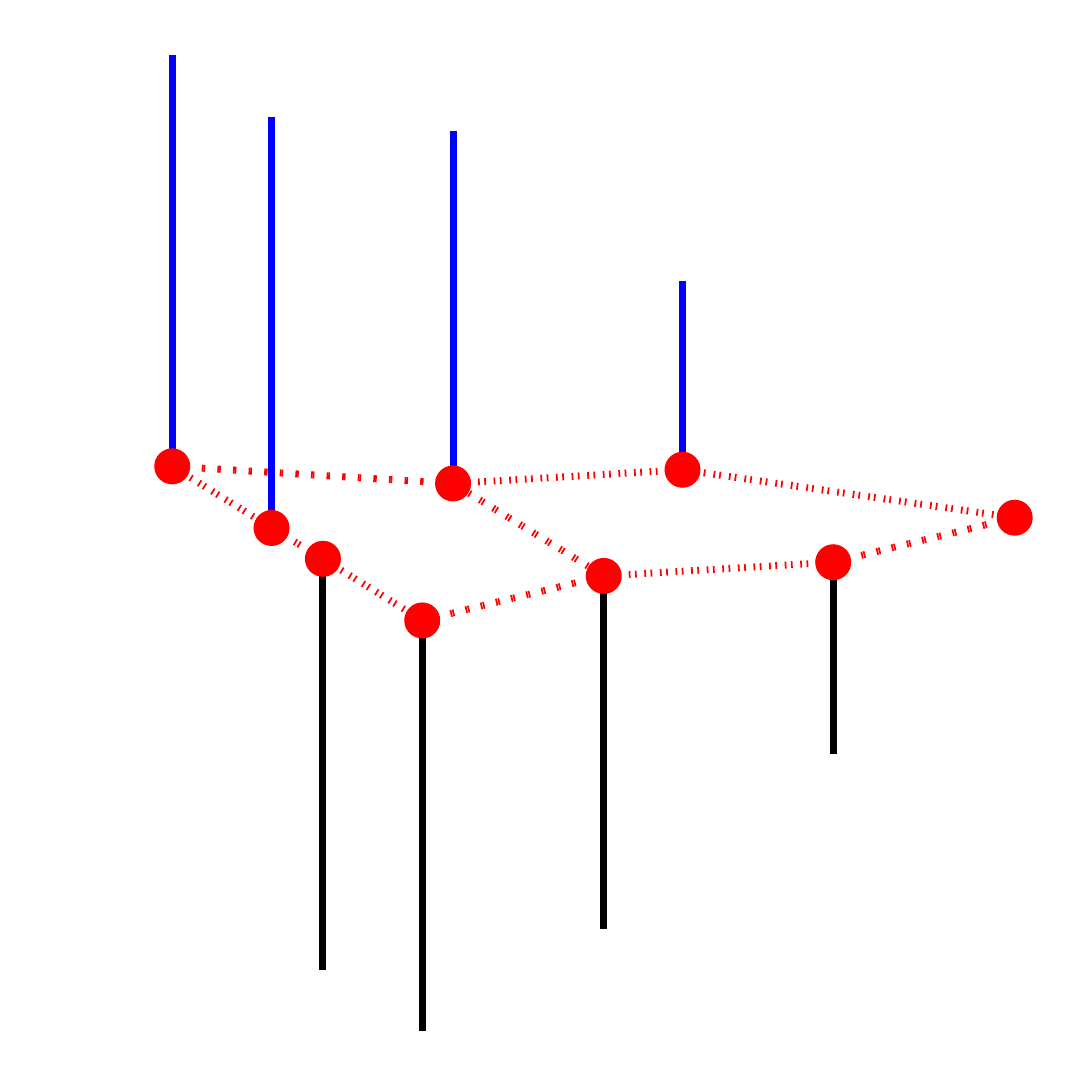}} 
   \centerline{\includegraphics[width=1.05\linewidth]{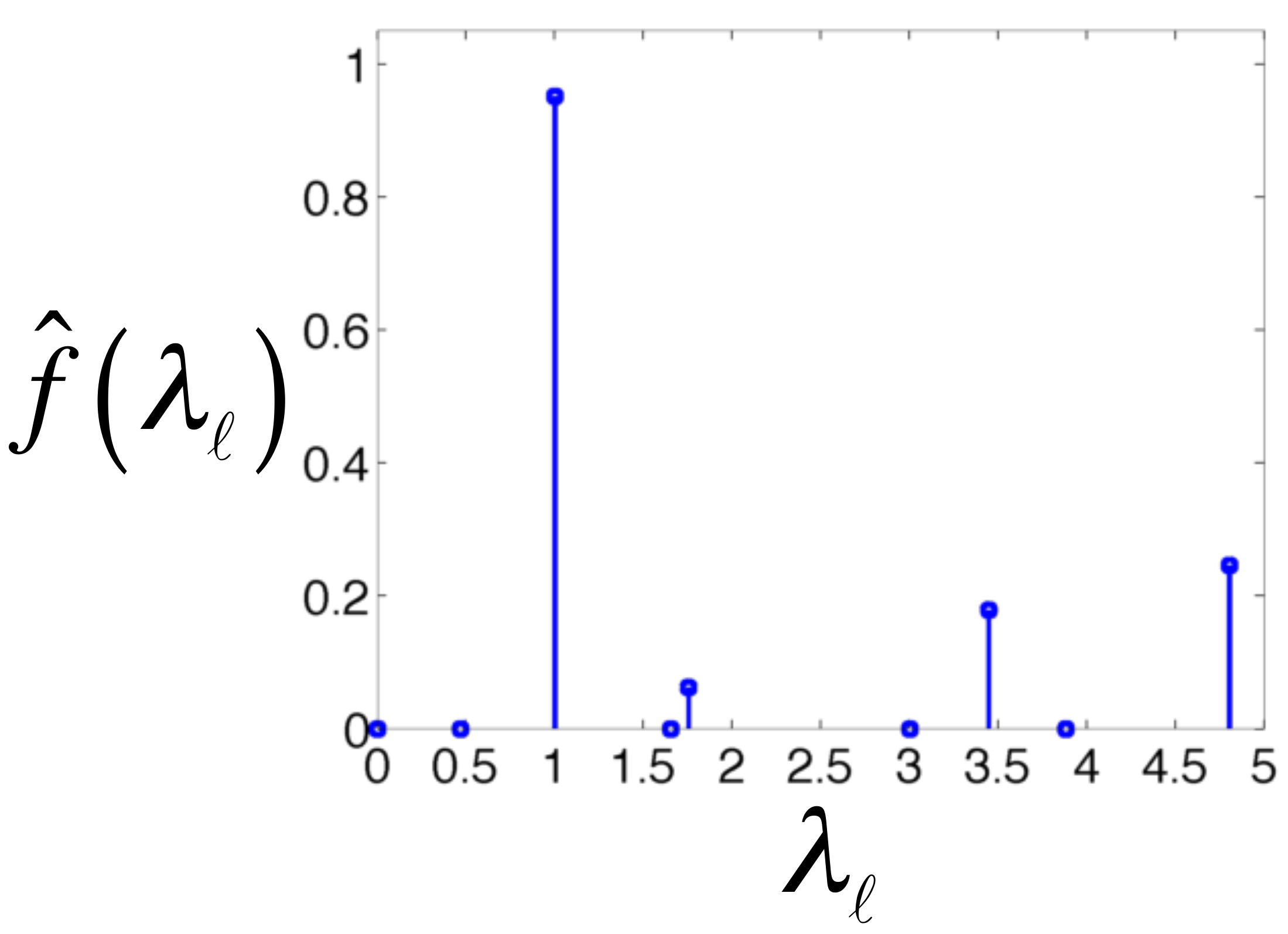}}

\end{minipage}
\hfill
\hfill
\begin{minipage}[b]{.30\linewidth}
   \centering
   \centerline{\small{~~~~$\G_3$}}
   \centerline{\includegraphics[width=\linewidth]{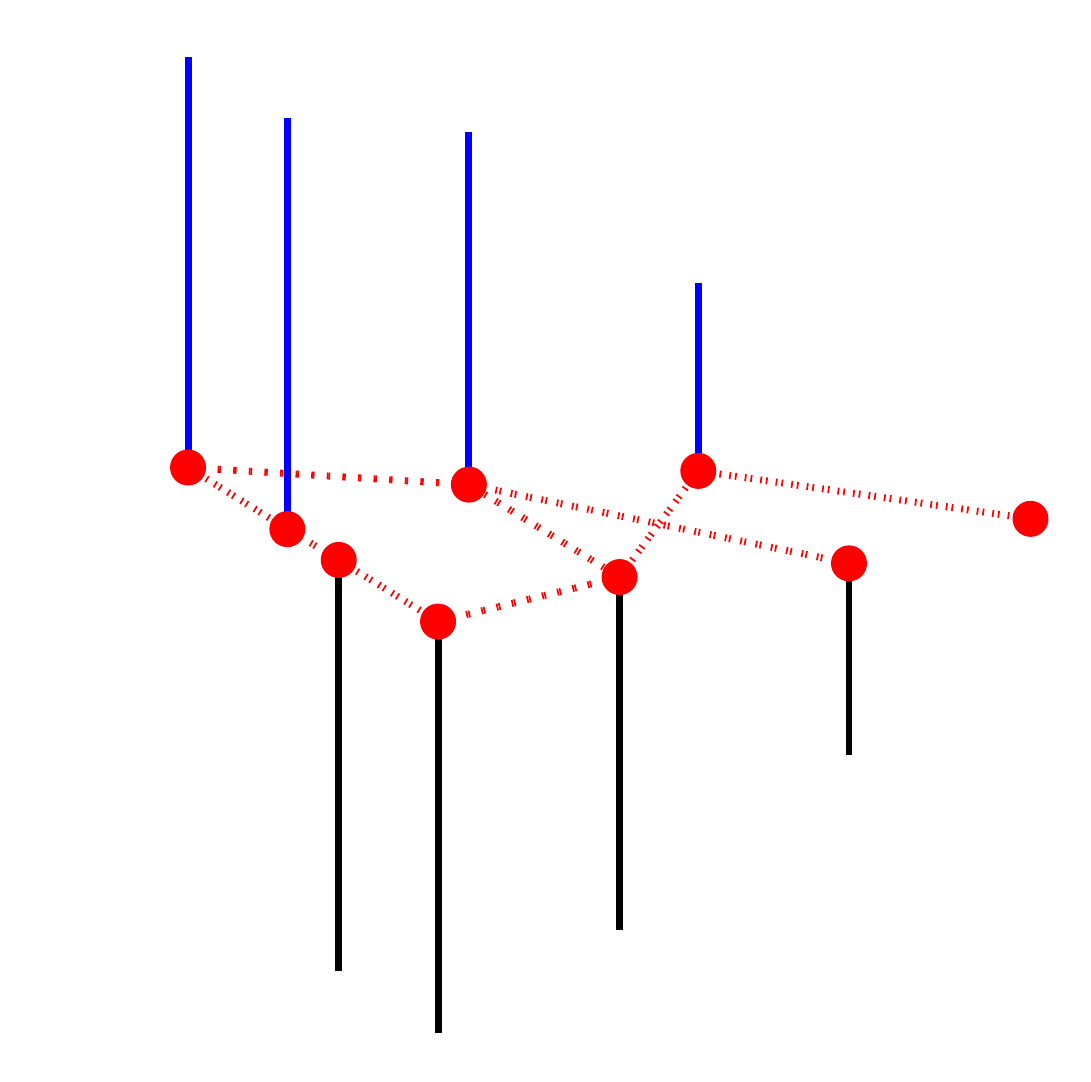}} 
   \centerline{\includegraphics[width=1.05\linewidth]{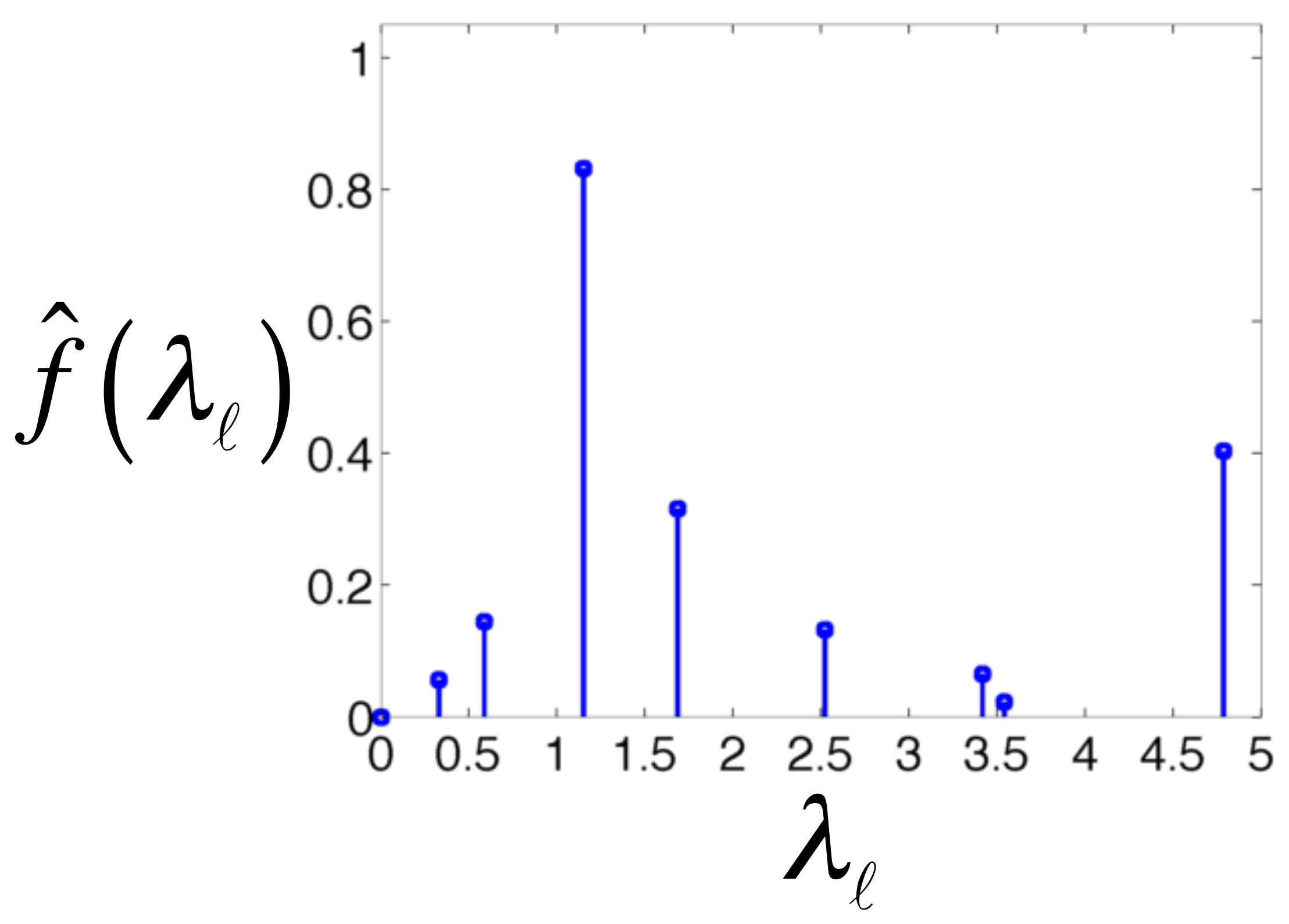}}
\end{minipage}
\hfill
\hfill 
\begin{example}[Importance of the underlying graph] \label{Ex:underlying}
In the figure above, we plot the same signal $\mathbf{f}$ on three different unweighted graphs 
with the same set of vertices, but different edges. The top row shows the signal in the vertex domains, and the bottom row shows the signal in the respective graph spectral domains. 

\hspace{.1in}The smoothness and graph spectral content of the signal both depend on the underlying graph structure. In particular, the signal $\mathbf{f}$ is smoothest with respect to the intrinsic structure of $\G_1$, and least smooth with respect to the intrinsic structure of $\G_3$. This can be seen (i) visually;
(ii) through the Laplacian quadratic form, as $\mathbf{f}^{\transpose}\L_1\mathbf{f}=0.14$, $\mathbf{f}^{\transpose}\L_2\mathbf{f}=1.31$, and $\mathbf{f}^{\transpose}\L_3\mathbf{f}=1.81$; and (iii) through the graph spectral representations, where the signal has all of its energy in the low frequencies in the graph spectral plot of $\hat{\mathbf{f}}$ on $\G_1$, and more energy in the higher frequencies in the graph spectral plot of $\hat{\mathbf{f}}$ on $\G_3$. 
\end{example}
\end{minipage}
}
\end{figure}

In summary, the connectivity of the underlying graph is encoded in the graph Laplacian, which is used to define both a graph Fourier transform (via the graph Laplacian eigenvectors) and different notions of smoothness. Example \ref{Ex:underlying} in the box below demonstrates how both the smoothness and the graph spectral content of a graph signal depend on the underlying graph.


\subsection{Other Graph Matrices} \label{Se:other_graph}

The basis $\{\mathbf{u}_{\l}\}_{\l=0,1,\ldots,N-1}$ of graph Laplacian eigenvectors is just one possible basis to use in the forward and inverse graph Fourier transforms \eqref{Eq:graph_FT} and \eqref{Eq:graph_IFT}. A second popular option is to normalize each weight $W_{i,j}$ by a factor of $\frac{1}{\sqrt{d_i d_j}}$. Doing so leads to the \emph{normalized graph Laplacian}, which is defined as 
$\tilde{\L}:=\mathbf{D}^{-\frac{1}{2}}\L\mathbf{D}^{-\frac{1}{2}}$, or, equivalently,
\begin{align*}
(\tilde{\L} f)(i)=\frac{1}{\sqrt{d_{i}}}\sum_{j \in \N_i} W_{i,j}\left[\frac{f(i)}{\sqrt{d_i}}-\frac{f(j)}{\sqrt{d_j}}\right].
\end{align*}
The eigenvalues $\{\tilde{\lambda}_{\l}\}_{\l=0,1,\ldots,N-1}$
of the normalized graph Laplacian of a connected graph $\G$ satisfy
\begin{align*}
0=\tilde{\lambda}_0 < \tilde{\lambda}_1 \leq \ldots \leq \tilde{\lambda}_{\max} \leq 2,
\end{align*}
with $\tilde{\lambda}_{\max}=2$ if and only if $\G$ is \emph{bipartite}; i.e., the set of vertices $\V$ can be partitioned into two subsets $\V_1$ and $\V_2$ such that every edge $e \in \E$ connects one vertex in $\V_1$ and one vertex in $\V_2$. We denote the normalized graph Laplacian eigenvectors by  $\{\tilde{\mathbf{u}}_{\l}\}_{\l=0,1,\ldots,N-1}$.
As seen in Figure \ref{Fig:zero_crossings}(b), the spectrum of $\tilde{\L}$ also carries a notion of frequency, with the eigenvectors associated with higher eigenvalues generally having more zero crossings. However, unlike $\mathbf{u}_0$, the normalized graph Laplacian eigenvector $\tilde{\mathbf{u}}_0$ associated with the zero eigenvalue is not a constant vector. 

The normalized and non-normalized graph Laplacians are both examples of \emph{generalized graph Laplacians} \cite[Section 1.6]{lap_eigen}, also called \emph{discrete Schr{\"o}dinger operators}. A generalized graph Laplacian of a graph $\G$ is any symmetric matrix whose $i,j^{th}$ entry is negative if there is an edge connecting vertices $i$ and $j$, equal to zero if $i \neq j$ and $i$ is not connected to $j$, and may be anything if $i=j$. 

A third popular matrix that is often used in dimensionality-reduction techniques for signals on graphs is the \emph{random walk matrix} $\mathbf{P}:=\mathbf{D}^{-1}\mathbf{W}$. Each entry $P_{i,j}$ describes the probability of going from vertex $i$ to vertex $j$ in one step of a Markov random walk on the graph $\G$. For connected, aperiodic graphs, each row of $\mathbf{P}^t$ converges to the stationary distribution of the random walk as $t$ goes to infinity. Closely related to the random walk matrix is the asymmetric graph Laplacian, which is defined as $\L_a:=\mathbf{I}_N-\mathbf{P}$, where $\mathbf{I}_N$ is the $N \times N$ identity matrix.\footnote{$\L_a$ is not a generalized graph Laplacian due to its asymmetry.} Note that $\L_a$ has the same set of eigenvalues as $\tilde{\L}$, and if $\tilde{\mathbf{u}}_{\l}$ is an eigenvector of $\tilde{L}$ associated with $\tilde{\lambda_{\l}}$, then $\mathbf{D}^{-\frac{1}{2}}\tilde{\mathbf{u}}_{\l}$ is an eigenvector of $\L_a$ associated with the eigenvalue $\tilde{\lambda_{\l}}$.

As discussed in detail in the next section, both the normalized and non-normalized graph Laplacian eigenvectors can be used as filtering bases. There is not a clear answer as to when to use the normalized graph Laplacian eigenvectors, when to use the non-normalized graph Laplacian eigenvectors, and when to use some other basis. The normalized graph Laplacian has the nice properties that its spectrum is always contained in the interval $[0,2]$ and, for bipartite graphs, the spectral folding phenomenon \cite{narang_bipartite_prod} can be exploited. However, the fact that the non-normalized graph Laplacian eigenvector associated with the zero eigenvalue is constant is a useful property in extending intuitions about DC components of signals from classical filtering theory. 

\section{Generalized Operators for Signals on Graphs} \label{Se:ops}

In this section, we review different ways to generalize fundamental operations such as filtering, translation, modulation, dilation, and downsampling to the graph setting. These generalized operators are the ingredients used to develop the localized, multiscale transforms described in
Section \ref{Se:transforms}.
 
\subsection{Filtering}
The first generalized operation we tackle is filtering. We start by extending the notion of frequency filtering to the graph setting, and then discuss localized filtering in the vertex domain. 
\subsubsection{Frequency Filtering}

In classical signal processing, frequency filtering is the process of representing an input signal as a linear combination of complex exponentials, and amplifying or attenuating the contributions of some of the component complex exponentials: 
\begin{align}\label{Eq:class_filter}
\hat{f}_{out}(\xi) = \hat{f}_{in}(\xi)\hat{h}(\xi),
\end{align}
where $\hat{h}(\cdot)$ is the transfer function of the filter. Taking an inverse Fourier transform of \eqref{Eq:class_filter}, multiplication in the Fourier domain corresponds to convolution in the time domain:
\begin{align} 
{f}_{out}(t)&=\int_{\Rbb}\hat{f}_{in}(\xi)\hat{h}(\xi) e^{2\pi i \xi t} d\xi  \label{Eq:conv_class} \\
&= \int_{\Rbb} f_{in}(\tau)h(t-\tau)d\tau=:(f_{in} \ast h)(t). \label{Eq:conv_class2}
\end{align}

Once we fix a graph spectral representation, and thus our notion of a 
graph Fourier transform (in this section, we use the eigenvectors of $\L$, but $\tilde{\L}$  
can also be used), we can directly generalize \eqref{Eq:class_filter} to define frequency filtering, or \emph{graph spectral filtering}, as
\begin{align} \label{Eq:filtering1}
\hat{f}_{out}(\lambda_{\l}) = \hat{f}_{in}(\lambda_{\l})\hat{h}(\lambda_{\l}),
\end{align}
or, equivalently, taking an inverse graph Fourier transform, 
\begin{align} \label{Eq:filtering2}
f_{out}(i) = \sum_{\l=0}^{N-1} \hat{f}_{in}(\lambda_{\l})\hat{h}(\lambda_{\l}) u_{\l}(i).
\end{align}
Borrowing notation from the theory of matrix functions \cite{higham}, we can also write \eqref{Eq:filtering1} and \eqref{Eq:filtering2} as $\mathbf{f}_{out}=\hat{h}(\L)\mathbf{f}_{in}$, where 
\begin{align}\label{Eq:filt_matrix}
\hat{h}(\L):=\mathbf{U}\left[
\begin{array}{ccc}
\hat{h}(\lambda_0) &&\mathbf{0} \\
&\ddots & \\
\mathbf{0}&&\hat{h}(\lambda_{N-1})
\end{array}
\right]\mathbf{U}^{\transpose}.
\end{align}

\begin{figure*}
\fboxsep=3mm
\fboxrule=2pt
\fcolorbox{darkkhaki}{lightkhaki}{
\begin{minipage}{6.75in}
\begin{example}[Tikhonov regularization] \label{Ex:tik_reg}
We observe a noisy graph signal $\mathbf{y}=\mathbf{f}_0 + \boldsymbol{\eta}$, where $\boldsymbol{\eta}$ is uncorrelated additive Gaussian noise, and we wish to recover $\mathbf{f}_0$. To enforce \emph{a priori} information that the clean signal $\mathbf{f}_0$ is smooth with respect to the underlying graph, we include a regularization term of the form $\mathbf{f}^{\transpose}\L \mathbf{f}$, and, for a fixed $\gamma>0$, solve the optimization problem
\begin{align}\label{Eq:reg_prob}
\argmin_{\mathbf{f}}\left\{\norm{\mathbf{f}-\mathbf{y}}_2^2+\gamma \mathbf{f}^{\transpose}\L\mathbf{f}\right\}.
\end{align}
The first-order optimality conditions of the convex objective function in \eqref{Eq:reg_prob} show that (see, e.g., \cite{smola}, \cite[Section III-A]{elmoataz},\cite[Proposition 1]{shuman_DCOSS_2011}) the optimal reconstruction is given by 
\begin{align}\label{Eq:tik_filtering}
f_*(i) = \sum_{\l=0}^{N-1} \left[\frac{1}{1+\gamma \lambda_{\l}}\right]\hat{y}(\lambda_{\l}) u_{\l}(i),
\end{align}
or, equivalently, $\mathbf{f}=\hat{h}(\L)\mathbf{y}$, where $\hat{h}(\lambda):=\frac{1}{1+\gamma \lambda}$ can be viewed as a low-pass filter. 

\hspace{.1in} 
As an example, in the figure below,  
we take the 512 x 512 cameraman image as $\mathbf{f}_0$ and corrupt it with additive Gaussian noise with mean zero and standard deviation 0.1 to get a noisy signal $\mathbf{y}$. We then apply two different filtering methods to denoise the signal. In the first method, we
apply a symmetric two-dimensional Gaussian low-pass filter of size 72 x 72 with two different standard deviations: 1.5 and 3.5.
 In the second method, we
 form a \emph{semi-local} graph on the pixels by connecting each pixel to its horizontal, vertical, and diagonal neighbors, and setting the Gaussian weights \eqref{Eq:gkw} between two neighboring pixels according to the similarity of the noisy image values at those two pixels; i.e., the edges of the semi-local graph are independent of the noisy image, but the distances in \eqref{Eq:gkw} are the differences between the neighboring pixel values in the noisy image. For the Gaussian weights in \eqref{Eq:gkw}, we take $\theta=0.1$ and $\kappa=0$.
 We then 
 perform the 
 low-pass graph filtering  \eqref{Eq:tik_filtering} with $\gamma=10$ to reconstruct the image. This method is a variant of the graph-based anisotropic diffusion image smoothing method of \cite{hancock}. 
 
 \hspace{.1in} In all image displays, we threshold the values to the [0,1] interval. The bottom row of images is comprised of zoomed-in versions of the top row of images. Comparing the results of the two filtering methods, we see that in order to smooth sufficiently in smoother areas of the image, the classical Gaussian filter also smooths across the image edges. The graph spectral filtering method does not smooth as much across the image edges, as the geometric structure of the image is encoded in the graph Laplacian via the noisy image.
\end{example}
\vspace{.15in}
\begin{minipage}[t]{.196\linewidth}
   \centering
   \centerline{\small{~}}
\end{minipage}
\begin{minipage}[t]{.196\linewidth}
   \centering
      \centerline{\small{~}}
\end{minipage}
\begin{minipage}[t]{.196\linewidth}
      \centerline{\small{Gaussian-Filtered}} 
\end{minipage}
\begin{minipage}[t]{.196\linewidth}
      \centerline{\small{~Gaussian-Filtered}} 
 \end{minipage}
\begin{minipage}[t]{.196\linewidth}
   \centering
      \centerline{\small{~}}
\end{minipage}
\\
\begin{minipage}[t]{.196\linewidth}
   \centering
   \centerline{\small{Original Image~~}}
\end{minipage}
\begin{minipage}[t]{.196\linewidth}
   \centering
      \centerline{\small{Noisy Image~}}
\end{minipage}
\begin{minipage}[t]{.196\linewidth}
      \centerline{\small{(Std. Dev. = 1.5)}} 
\end{minipage}
\begin{minipage}[t]{.196\linewidth}
      \centerline{\small{~(Std. Dev. = 3.5)}} 
\end{minipage}
\begin{minipage}[t]{.196\linewidth}
   \centering
      \centerline{\small{~~~Graph-Filtered}}
\end{minipage}
\\
\hfill
\begin{minipage}[t]{\linewidth}
   \centering
   \centerline{\includegraphics[width=6.5in]{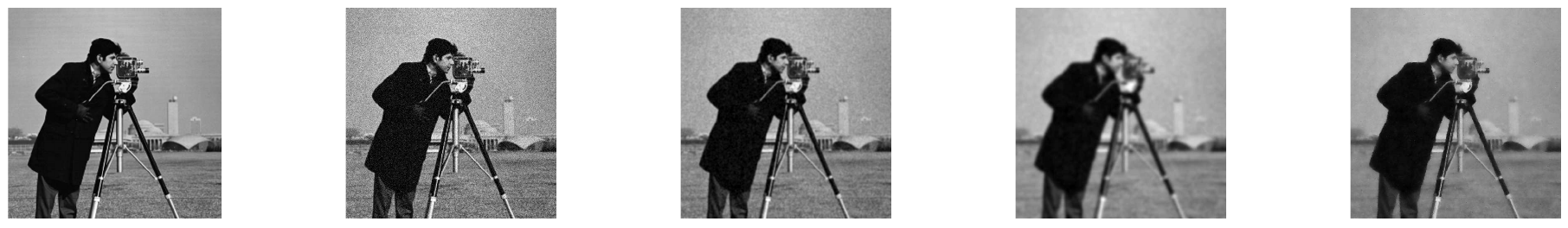}} 
\end{minipage}
\hfill 
\\
\hfill
\begin{minipage}[t]{\linewidth}
   \centering
   \centerline{\includegraphics[width=6.5in]{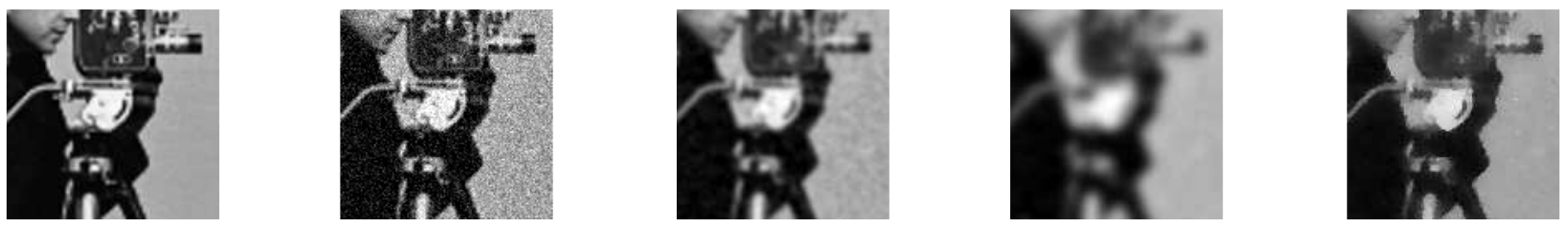}} 
\end{minipage}
\hfill 
\end{minipage}
}

\end{figure*}

The basic graph spectral filtering \eqref{Eq:filtering1} can be used to implement discrete versions of well-known continuous filtering techniques such as Gaussian smoothing, bilateral filtering, total variation filtering, anisotropic diffusion, and non-local means filtering (see, e.g., \cite{buades} and references therein). In particular, many of these filters arise as solutions to 
variational problems to regularize ill-posed inverse problems such as denoising, inpainting, and super-resolution.  One example is the discrete regularization framework
\begin{align}\label{Eq:energy_min}
\min_{\mathbf{f}}\left\{\norm{\mathbf{f}-\mathbf{y}}_2^2+\gamma S_p(\mathbf{f})\right\},
\end{align}
where $S_p(\mathbf{f})$ is the p-Dirichlet form of \eqref{Eq:pdirichlet}. References
\cite{smola,harmonic,zhou_scholkopf,reg_discrete,belkin_matveeva,zhou_bousquet,hancock,peyre_nlr}, \cite[Chapter 5]{grady}, and \cite{elmoataz,bougleux,osher_chapter} discuss \eqref{Eq:energy_min} and other energy minimization models in detail, as well as specific filters that arise as solutions, relations between these discrete graph spectral filters and filters arising out of continuous partial differential equations, and applications such as graph-based image processing, mesh smoothing, and statistical learning. In Example \ref{Ex:tik_reg}, we show one particular image denoising application of \eqref{Eq:energy_min} with $p=2$.

\subsubsection{Filtering in the Vertex Domain}
To filter a signal in the vertex domain, we simply write the output $f_{out}(i)$ 
 at vertex $i$ as a linear combination of the components of the input signal at vertices within a $K$-hop local neighborhood of vertex $i$:
\begin{align}\label{Eq:vertex_filtering}
f_{out}(i) = b_{i,i} f_{in}(i) + \sum_{j \in \N(i,K)} b_{i,j} f_{in}(j),
\end{align}
for some constants $\{b_{i,j}\}_{i,j\in\V}$. Equation \eqref{Eq:vertex_filtering} just says that filtering in the vertex domain is a localized linear transform.

We now briefly relate filtering in the graph spectral domain (frequency filtering) to filtering in the vertex domain. 
When the frequency filter in \eqref{Eq:filtering1} is an order $K$ polynomial $\hat{h}(\lambda_{\l})=\sum_{k=0}^K a_k \lambda_{\l}^k$ for some constants $\{a_k\}_{k=0,1,\ldots,K}$, we can also interpret the filtering equation \eqref{Eq:filtering1} in the vertex domain. From \eqref{Eq:filtering2}, we have
\begin{align} \label{Eq:filtering_vertex1}
f_{out}(i) &= \sum_{\l=0}^{N-1} \hat{f}_{in}(\lambda_{\l})\hat{h}(\lambda_{\l}) u_{\l}(i) \nonumber \\
&=\sum_{j=1}^N f_{in}(j) \sum_{k=0}^K a_k \sum_{\l=0}^{N-1}\lambda_{\l}^k u_{\l}^*(j) u_{\l}(i) \nonumber \\
&= \sum_{j=1}^N f_{in}(j) \sum_{k=0}^K a_k \left(\L^k\right)_{i,j}.
\end{align}
Yet, $\left(\L^k\right)_{i,j}=0$ when the shortest-path distance $d_{\G}(i,j)$ between vertices $i$ and $j$ (i.e. the minimum number of edges comprising any path connecting $i$ and $j$) is greater than $k$ \cite[Lemma 5.2]{sgwt}. Therefore, we can write \eqref{Eq:filtering_vertex1} exactly as in \eqref{Eq:vertex_filtering}, with
the constants 
defined as 
\begin{align*}
b_{i,j}:=\sum_{k=d_{\G}(i,j)}^K a_k \left(\L^k\right)_{i,j}.
\end{align*}
So when the frequency filter is an order $K$ polynomial, the frequency filtered signal at vertex $i$, $f_{out}(i)$, is a linear combination of the components of the input signal at vertices within a $K$-hop local neighborhood of vertex $i$. This property can be quite useful when relating the smoothness of a filtering kernel to the localization of filtered signals in the vertex domain.

\subsection{Convolution}
We cannot directly generalize the definition \eqref{Eq:conv_class2} of a convolution product to the graph setting, because of the 
term $h(t-\tau)$.
However, 
one way to define a generalized convolution product for signals on graphs is to replace 
the complex exponentials in \eqref{Eq:conv_class} with the graph Laplacian eigenvectors \cite{shuman_SSP_2012}:
\begin{align}\label{Eq:gen_convolution}
(f \ast h)(i):=\sum_{\l=0}^{N-1} \hat{f}(\lambda_{\l})\hat{h}(\lambda_{\l}) u_{\l}(i),
\end{align}
which enforces the property that convolution in the vertex domain is equivalent to multiplication in the graph spectral domain.

\subsection{Translation}
\begin{figure}[h]
\hfill
\begin{minipage}[b]{.32\linewidth}
   \centering
   \centerline{\includegraphics[width=\linewidth]{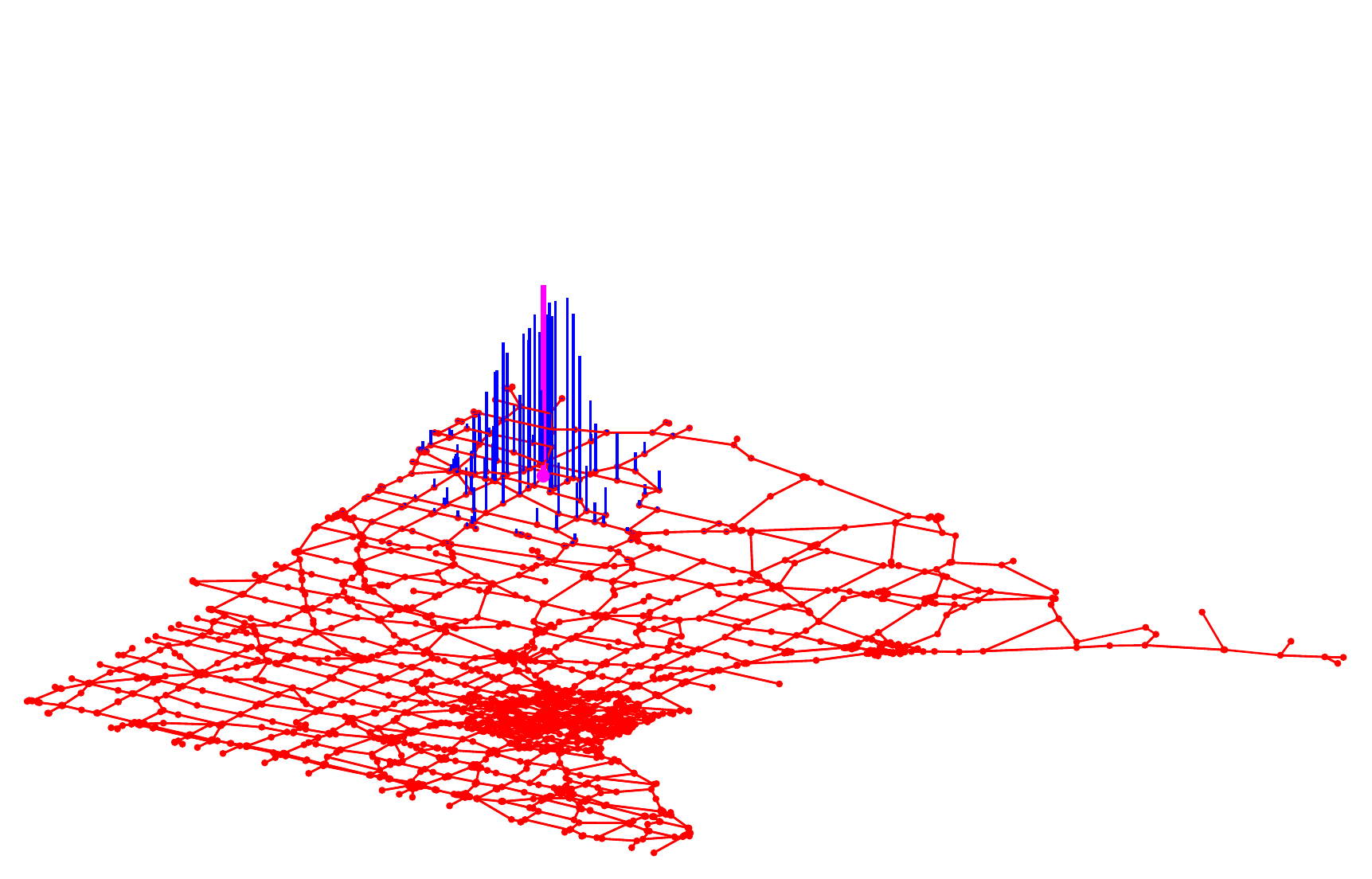}} 
   \centerline{\small{(a)}}
\end{minipage}
\hfill
\begin{minipage}[b]{.32\linewidth}
   \centering
      \centerline{\includegraphics[width=\linewidth]{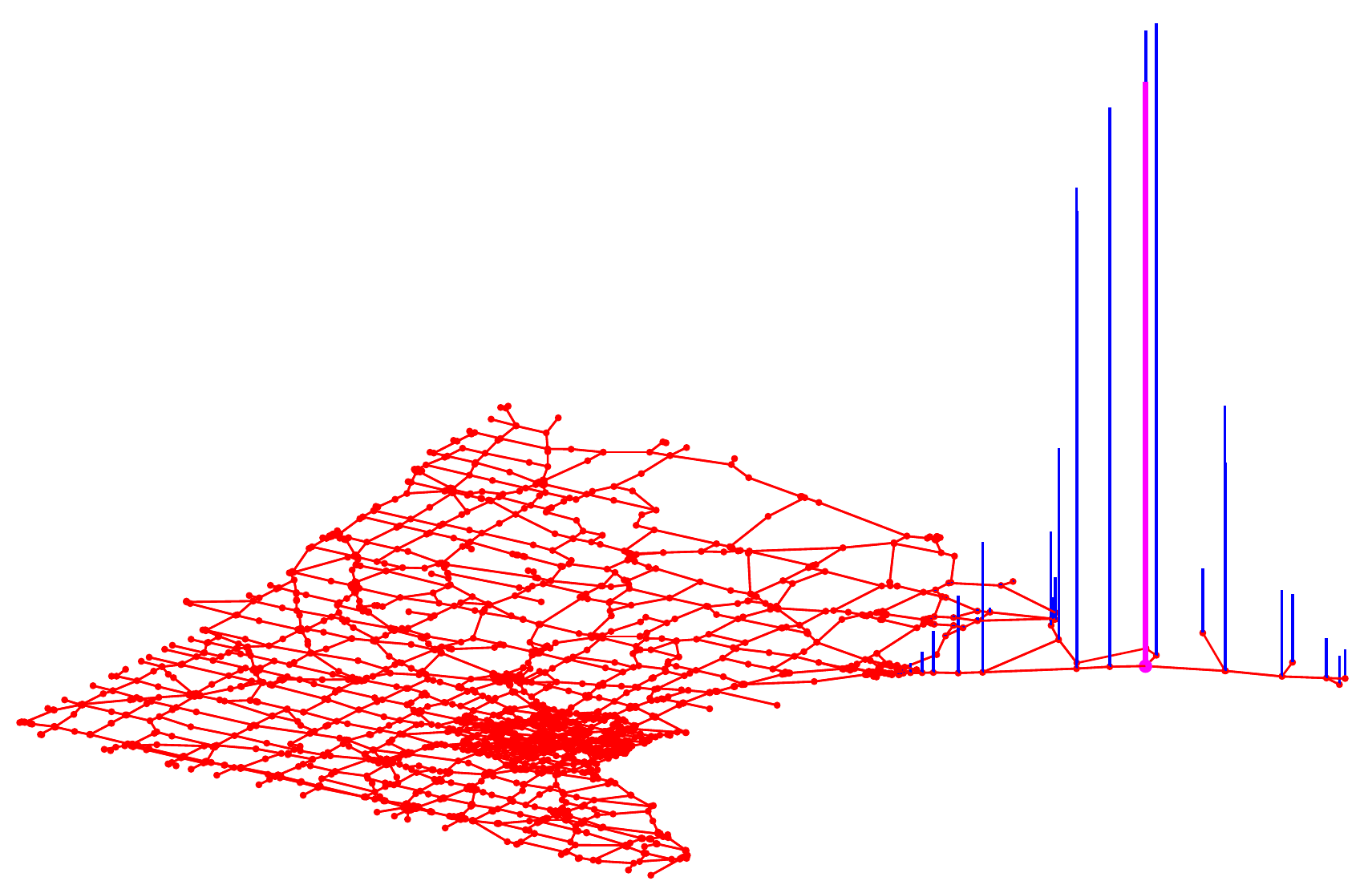}} 
      \centerline{\small{(b)}}
\end{minipage}
\hfill
\hfill
\begin{minipage}[b]{.32\linewidth}
   \centering
   \centerline{\includegraphics[width=\linewidth]{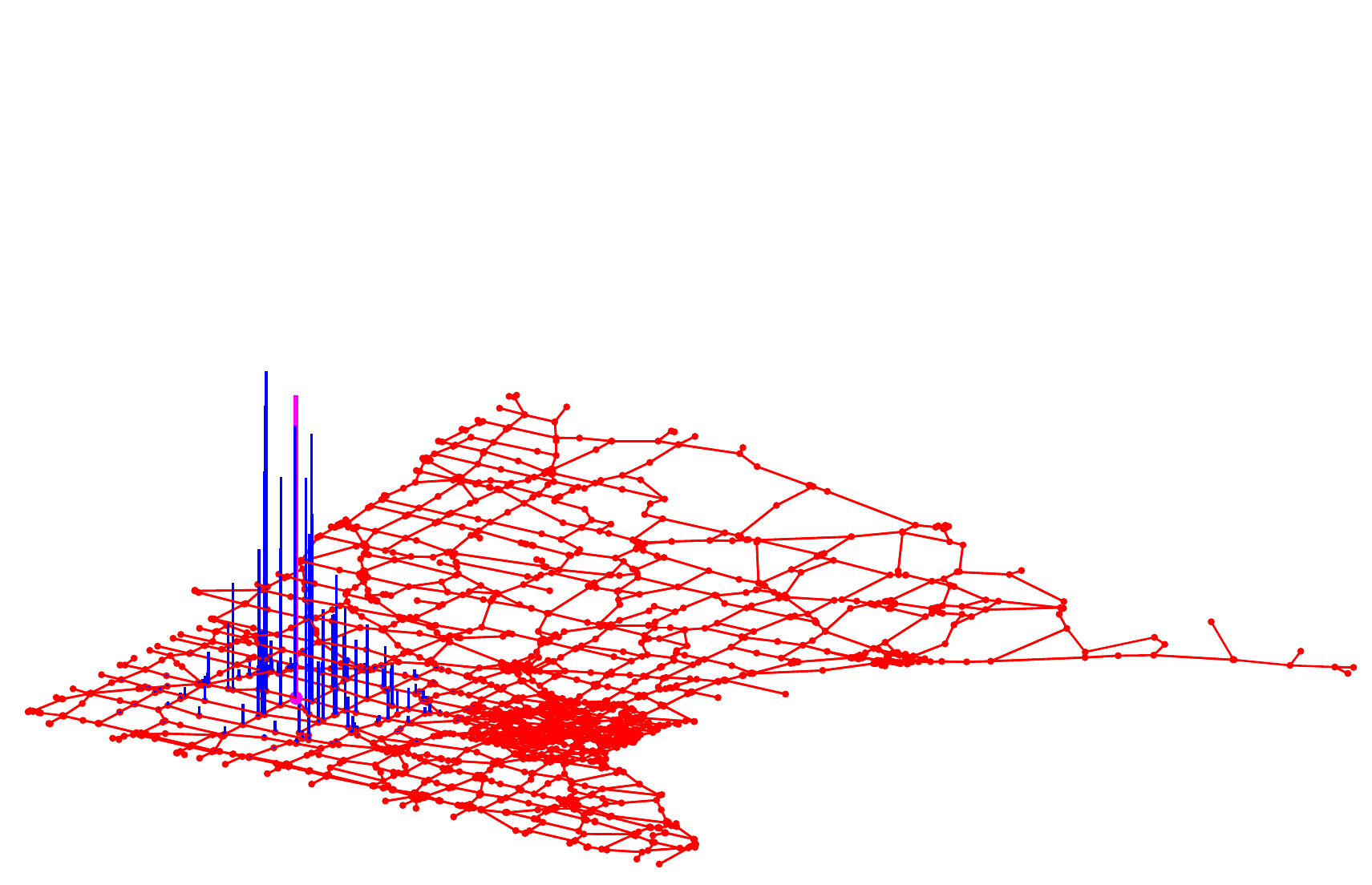}}
   \centerline{\small{(c)}}
\end{minipage}
\hfill
\hfill
\caption {The translated signals (a) $T_{100}\mathbf{g}$, (b) $T_{200}\mathbf{g}$, and (c) $T_{2000}\mathbf{g}$, where $\mathbf{g}$ is the heat kernel shown in Figures \ref{Fig:minn_signals}(a) and \ref{Fig:minn_signals}(b).}
  \label{Fig:translation}
\end{figure}

The classical translation operator is defined through the change of variable $(T_{\upsilon} f)(t):=f(t-{\upsilon})$, which, as discussed earlier, we cannot directly generalize to the graph setting. However, we can also view the classical translation operator $T_{\upsilon}$ as a convolution with a delta centered at ${\upsilon}$; i.e., $(T_{\upsilon} f)(t)=(f \ast \delta_{\upsilon})(t)$ in the weak sense. 
Thus, one way to define a \emph{generalized translation operator} $T_{n}: \Rbb^N \rightarrow \Rbb^N$ is via generalized convolution with a delta centered at vertex $n$ \cite{sgwt,shuman_SSP_2012}: 
\begin{align} \label{Eq:new_translation}
\left(T_n g\right)(i):= \sqrt{N}(g \ast \delta_n)(i) \stackrel{\eqref{Eq:gen_convolution}}= 
\sqrt{N}\sum_{\l=0}^{N-1}\hat{g}(\lambda_{\l})u_{\l}^*(n)u_{\l}(i),
\end{align}
where 
\begin{align}
\delta_{n}(i)=
\begin{cases}
1 &\mbox{if } i=n \\
0 &\mbox{otherwise}
\end{cases}.
\end{align}
A few remarks about the generalized translation \eqref{Eq:new_translation} are in order. 
First, we do not usually view it as translating a signal $\mathbf{g}$ defined in the vertex domain, but rather as a kernelized operator acting on a kernel $\hat{{g}}(\cdot)$ defined directly in the graph spectral domain. To translate this kernel to vertex $n$, the $\l^{th}$ component of the kernel is multiplied by $u_{\l}^*(n)$, and then an inverse graph Fourier transform is applied. 
Second, the normalizing constant $\sqrt{N}$ in \eqref{Eq:new_translation} ensures that the translation operator preserves the mean of a signal; i.e., $\sum_{i=1}^N(T_n g)(i)=\sum_{i=1}^N g(i)$. Third, the smoothness of the kernel $\hat{g}(\cdot)$ controls the localization of $T_n \mathbf{g}$ around the center vertex $n$; that is, the magnitude $(T_n g)(i)$ of the translated kernel at vertex $i$ decays as the distance between $i$ and $n$ increases \cite{sgwt}. 
This property can be seen in Figure \ref{Fig:translation}, where we translate a heat kernel around to different locations of the Minnesota graph. Finally, unlike the classical translation operator, the generalized translation operator \eqref{Eq:new_translation} is not generally an isometric operator ($\norm{T_n \mathbf{g}}_2 \neq \norm{\mathbf{g}}_2$), due to the possible localization of the graph Laplacian eigenvectors ($\mu > \frac{1}{\sqrt{N}})$. 

\subsection{Modulation and Dilation} 
In addition to translation, many classical transform methods rely on modulation or dilation to localize signals' frequency content. The classical modulation operator
\begin{align} \label{Eq:class_mod}
(M_{\omega} f)(t):=e^{2\pi i \omega t} f(t)
\end{align} 
represents a translation in the Fourier domain:
\begin{align*}
\widehat{M_{\omega}f}(\xi)=\hat{f}(\xi-\omega),~\forall \xi \in \Rbb.
\end{align*}
One way to define generalized modulation in the graph setting is to replace the multiplication by a complex exponential (an eigenfunction of the 1D Laplacian operator) in \eqref{Eq:class_mod} with a multiplication by a graph Laplacian eigenvector: 
\begin{align}\label{Eq:gen_mod}
\left(M_{k}g\right)(i):=\sqrt{N}u_{k}(i)g(i).
\end{align}
The generalized modulation \eqref{Eq:gen_mod} is not exactly a translation in the graph spectral domain due to the discrete and irregular nature of the spectrum; however, as shown in \cite[Figure 3]{shuman_SSP_2012}, if a 
kernel $\hat{g}(\cdot)$ is localized around 0 in the graph spectral domain, then $\widehat{M_k \mathbf{g}}$ is localized around $\lambda_k$. 

For $s > 0$, dilation or scaling of an analog signal $f$ in the time domain is given by
\begin{align}\label{Eq:class_dil}
(\D_s f)(t):=\frac{1}{s}f \left(\frac{t}{s}\right).
\end{align}
We cannot directly generalize \eqref{Eq:class_dil} to the graph setting, because $\frac{i}{s}$ is not likely to be in the domain $\V$ for all $i \in \V$. Instead, we can take the Fourier transform of \eqref{Eq:class_dil}
\begin{align}\label{Eq:class_dil_four}
(\widehat{\D_s f})(\xi)=\hat{f}(s \xi),
\end{align}
and generalize \eqref{Eq:class_dil_four} to the graph setting. Assuming we start with a kernel $\hat{g}: \Rbb_+ \rightarrow \Rbb$, we can define a generalized graph dilation by \cite{sgwt}
\begin{align}\label{Eq:gen_dil}
(\widehat{\D_s{g}})(\lambda):=\hat{g}(s\lambda).
\end{align}
Note that, unlike the generalized modulation \eqref{Eq:gen_mod}, the generalized dilation \eqref{Eq:gen_dil} requires the kernel $\hat{g}(\cdot)$ to be defined on the entire real line, not just on $\sigma(\L)$ or $[0,\lambda_{\max}]$.

\begin{figure*}
\fboxsep=3mm
\fboxrule=2pt
\fcolorbox{darkkhaki}{lightkhaki}{
\begin{minipage}{6.75in}
\begin{example}[Diffusion operators and dilation] \label{Ex:diffusion}
The heat diffusion operator $\mathbf{R}=e^{-\L}$ is an example of a discrete diffusion operator 
(see, e.g., \cite{coifman_pnas} and \cite[Section 2.5.5]{grady} for general discussions of discrete diffusions and \cite[Section 4.1]
{diffusion_wavelets} for a formal definition and examples of symmetric diffusion semigroups). Intuitively, applying different powers $\tau$ of the heat diffusion operator to a signal $\mathbf{f}$ describes the flow of heat over the graph when the rates of flow are proportional to the edge weights encoded in $\L$. 
The signal $\mathbf{f}$ represents the initial amount of heat at each vertex, and $\mathbf{R}^\tau \mathbf{f}=\left(e^{-\tau\L}\right)\mathbf{f}$ represents the amount of heat at each vertex after time $\tau$. The time variable $\tau$ also provides a notion of scale. When $\tau$ is small, the entry $\left(e^{-\tau\L}\right)_{i,j}$ for two vertices that are far apart in the graph is very small, and therefore $\left(\left(e^{-\tau\L}\right)\mathbf{f}\right)(i)$ depends primarily on the values $f(j)$ for vertices $j$ close to $i$ in the graph. As $\tau$ increases, $\left(\left(e^{-\tau\L}\right)\mathbf{f}\right)(i)$ also depends on the values $f(j)$ for vertices $j$ farther away from $i$ in the graph. Zhang and Hancock \cite{hancock} provide a more detailed mathematical justification behind this migration from domination of the local geometric structures to domination of the global structure of the graph as $\tau$ increases, as well as a nice illustration of heat diffusion on a graph in  \cite[Figure 1]{hancock}.

\hspace{.1in} Using our notations from \eqref{Eq:filt_matrix} and \eqref{Eq:gen_dil}, we can see that applying a power $\tau$ of the heat diffusion operator to any signal $\mathbf{f}\in\Rbb^N$ is equivalent to filtering the signal with a dilated heat kernel:
\begin{align*}
\mathbf{R}^{\tau} \mathbf{f} = \left({e^{-\tau \L}}\right) \mathbf{f} = \widehat{\left(\D_{\tau}g\right)}(\L)\mathbf{f}=\mathbf{f} \ast \left(\D_{\tau}\mathbf{g}\right),
\end{align*}
where the filter is the heat kernel $\hat{g}(\lambda_{\l})=e^{-\lambda_{\l}}$, similar to the one shown in Figure \ref{Fig:minn_signals}(b). 

\hspace{.1in} 
In the figure below, we consider the cerebral cortex graph described in \cite{sgwt}, initialize a unit of energy at the vertex 100 by taking $\mathbf{f}=\boldsymbol{\delta}_{100}$, allow it to diffuse through the network for different dyadic amounts of time, and measure the amount of energy that accumulates at each vertex. Applying different powers of the heat diffusion operator can be interpreted as graph spectral filtering with a dilated kernel. The original signal $\mathbf{f}=\boldsymbol{\delta}_{100}$ on the cerebral cortex graph is shown in (a); the filtered signals 
$\left\{\mathbf{f}\ast \left(\D_{2^k-1}\mathbf{g}\right)\right\}_{k=1,2,3,4}=
\left\{\mathbf{R}^{2^k-1}\mathbf{f}\right\}_{k=1,2,3,4}$ are shown in (b)-(e); and the different dilated kernels corresponding to the dyadic powers of the diffusion operator are shown in (f).
Note that dyadic powers of diffusion operators of the form $\{\mathbf{R}^{2^k-1}\}_{k=1,2,\ldots}$ are of central importance to diffusion wavelets and diffusion wavelet packets \cite{diffusion_wavelets, Maggioni_biorthogonal,bremer_packets}, which we discuss in Section \ref{Se:transforms}.
\end{example}
\vspace{.15in}
\hfill
\begin{minipage}[b]{.15\linewidth}
   \centering
      \centerline{\small{$\boldsymbol{\delta}_{100}$~~}}
         \centerline{\includegraphics[width=\linewidth]{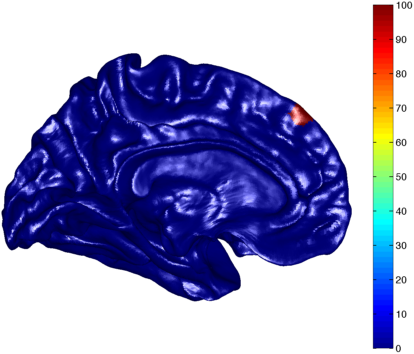}} 
   \centerline{\small{(a)}}
\end{minipage}
\hfill
\begin{minipage}[b]{.15\linewidth}
   \centering
         \centerline{\small{$e^{-\L}\boldsymbol{\delta}_{100}$~~}}
           \centerline{\includegraphics[width=\linewidth]{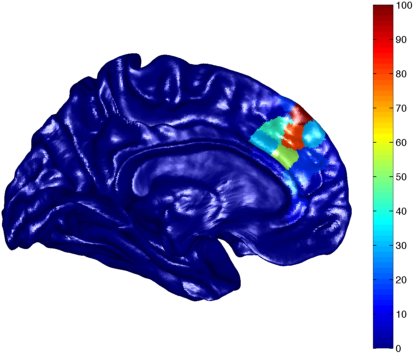}} 
      \centerline{\small{(b)}}
\end{minipage}
\hfill
\begin{minipage}[b]{.15\linewidth}
   \centering
            \centerline{\small{$e^{-3\L}\boldsymbol{\delta}_{100}$~~}}  \vspace{.05cm}
   \centerline{\includegraphics[width=\linewidth]{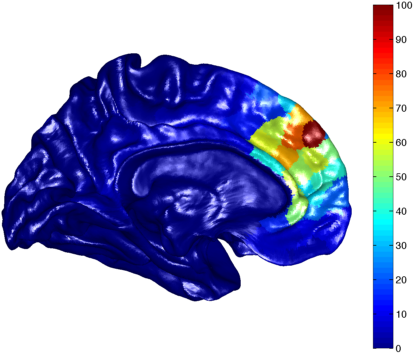}} 
   \centerline{\small{(c)}}
\end{minipage}
\hfill
\begin{minipage}[b]{.15\linewidth}
   \centering
            \centerline{\small{$e^{-7\L}\boldsymbol{\delta}_{100}$~~}}
              \centerline{\includegraphics[width=\linewidth]{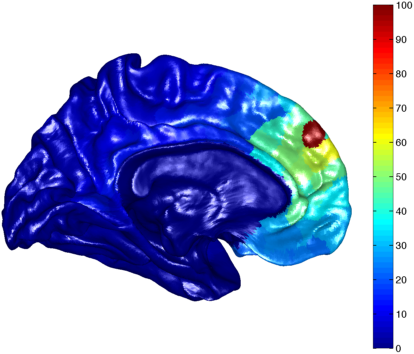}} 
   \centerline{\small{(d)}}
\end{minipage}
\hfill
\begin{minipage}[b]{.15\linewidth}
   \centering
            \centerline{\small{$e^{-15\L}\boldsymbol{\delta}_{100}$~~}}
   \centerline{\includegraphics[width=\linewidth]{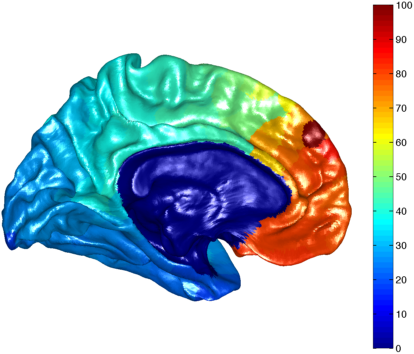}} 
      \centerline{\small{(e)}}
\end{minipage}
\hfill
\begin{minipage}[b]{.21\linewidth}
   \centering
   \centerline{\includegraphics[width=\linewidth]{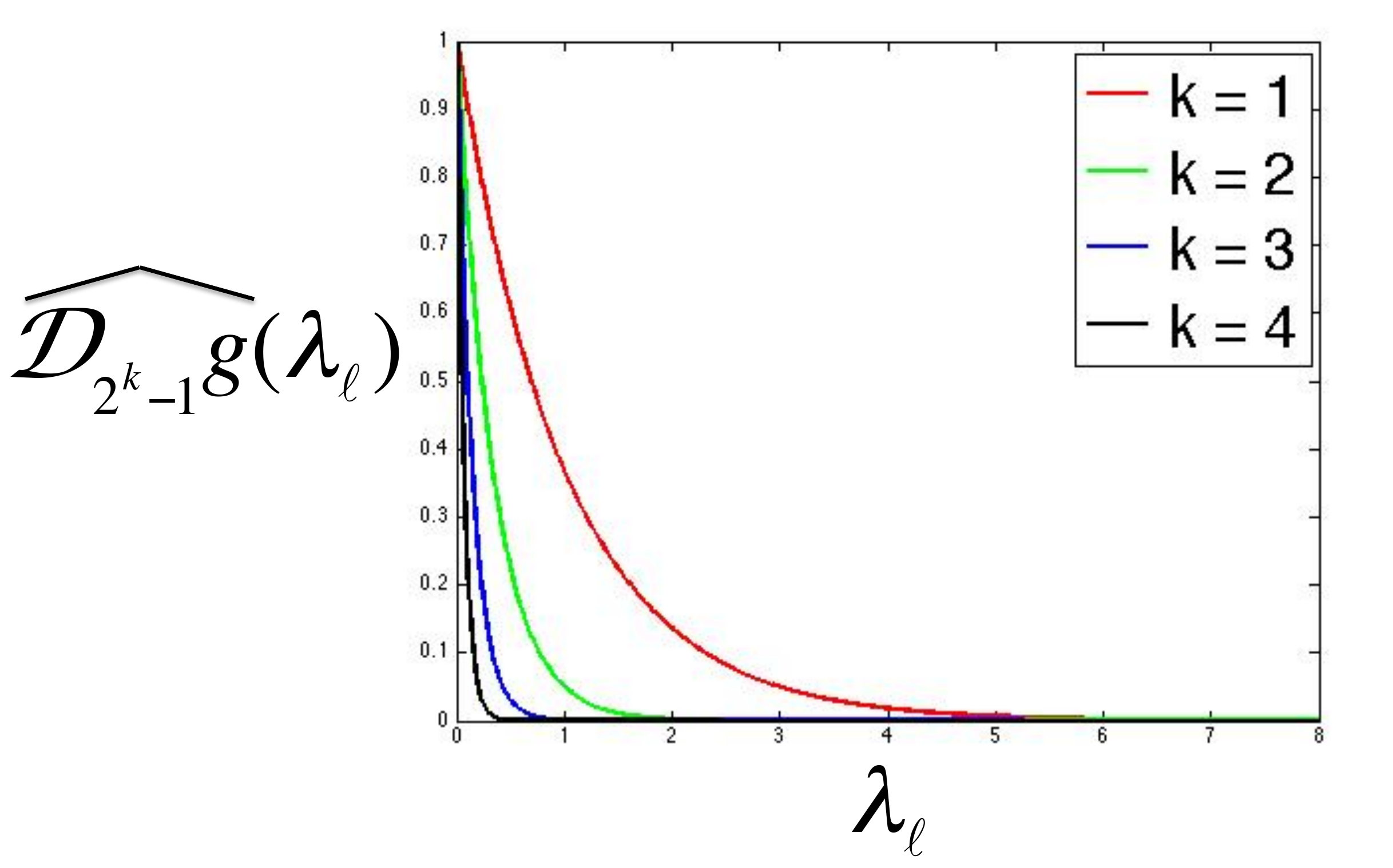}}
   \centerline{\small{~~(f)}}
\end{minipage}
\hfill
\hfill 
\end{minipage}
}
\end{figure*}

\subsection{Graph Coarsening, Downsampling, and Reduction}
Many multiscale transforms for signals on graphs require successively coarser versions of the original graph that preserve properties of the original graph such as the intrinsic geometric structure (e.g., some notion of distance between vertices), connectivity, graph spectral distribution, and sparsity. The process of transforming a given (fine scale) graph $\G=\{\V,\E,\mathbf{W}\}$ into a coarser graph $\G^{reduced}=\{\V^{reduced},\E^{reduced},\mathbf{W}^{reduced}\}$ with fewer vertices and edges, while also preserving the aforementioned properties, is often referred to as \emph{graph coarsening} or \emph{coarse-graining}\cite{lafon_coarse}.  

This process can be split into two separate but closely related subtasks: 1) identifying a reduced set of vertices $\V^{reduced}$, and 2) assigning edges and weights, $\E^{reduced}$ and $\mathbf{W}^{reduced}$, to connect the new set of vertices. When an additional constraint that $\V^{reduced} \subset \V$ is imposed, the first subtask is often referred to as \emph{graph downsampling}. The second subtask is often referred to as \emph{graph reduction} or \emph{graph contraction}.

In the special case of a bipartite graph, two subsets can be chosen so that every edge connects vertices in two different subsets. Thus, for bipartite graphs, there is a natural way to downsample by a factor of two, as there exists a notion of ``every other vertex.'' 

For non-bipartite graphs, the situation is far more complex, and a wide range of interesting techniques for the graph coarsening problem have been proposed by graph theorists, and, in particular, by the numerical linear algebra community.
To mention just a few, Lafon and Lee \cite{lafon_coarse} downsample based on \emph{diffusion distances} and form new edge weights based on random walk transition probabilities; the greedy \emph{seed selection} algorithm of Ron et al. \cite{ron} leverages an \emph{algebraic distance} measure to downsample the vertices; recursive spectral bisection \cite{barnard} repeatedly divides the graph into parts according to the polarity (signs) of the Fiedler vectors $\mathbf{u}_1$ of successive subgraphs; Narang and Ortega \cite{narang_lifting_graphs} minimize the number of edges connecting two vertices in the same downsampled subset; and another generally-applicable method which yields the natural downsampling on bipartite graphs (\cite[Chapter 3.6]{lap_eigen}) is to partition $\V$ into two subsets according to the polarity of the components of the graph Laplacian eigenvector $\mathbf{u}_{N-1}$ associated with the largest eigenvalue $\lambda_{\max}$.  We refer readers to \cite{ron,gp_archive} and references therein for more thorough reviews of the graph coarsening literature. 

There are also many interesting connections between graph coarsening, graph coloring \cite{aspvall}, spectral clustering \cite{spectral_clustering}, and nodal domain theory \cite[Chapter 3]{lap_eigen}. Finally, in a closely related topic, Pesenson (e.g., \cite{pesenson_paley}) has extended the concept of bandlimited sampling to signals defined on graphs by showing that certain classes of signals can be downsampled on particular subgraphs and then stably reconstructed from the reduced set of samples.

\section{Localized, Multiscale Transforms for Signals on Graphs} \label{Se:transforms}

The increasing prevalence of signals on graphs has triggered a recent influx of localized transform methods specifically designed to analyze 
data on graphs. 
These include
wavelets on unweighted graphs
for analyzing computer network traffic \cite{Crovella2003},
diffusion wavelets and diffusion wavelet packets
\cite{diffusion_wavelets, Maggioni_biorthogonal,bremer_packets}, the ``top-down'' wavelet construction of \cite{szlam}, graph dependent basis functions for sensor network graphs \cite{wang}, lifting based wavelets on graphs \cite{jansen, narang_lifting_graphs}, multiscale wavelets on balanced trees \cite{gavish},   spectral graph wavelets \cite{sgwt}, critically-sampled two-channel wavelet filter banks \cite{narang_icip, narang_bipartite_prod}, and
a windowed graph Fourier transform \cite{shuman_SSP_2012}.

Most of these designs 
are generalizations of the classical wavelet filter banks used to analyze signals on Euclidean domains. 
The feature that makes the classical 
wavelet transforms so 
useful 
is their ability to simultaneously
localize signal information in both 
time (or space) and frequency,
and thus   
exploit the time-frequency resolution
trade-off better than the 
Fourier transform. 
In a similar vein, the desired property
of 
wavelet transforms on graphs  
is to localize graph signal contents 
in both the vertex and graph spectral domains.  
In the classical setting, locality is measured in terms of the 
``spread'' of the signal 
in time and frequency, and 
 {\em uncertainty principles} 
 (see~ \cite[Sec. 2.6.2]{Vetterli_book}) 
describe 
the trade-off between time and frequency resolution.
Whether such a trade-off 
exists for graph signals 
remains an open question.
However, some recent works have begun to define different ways to measure the ``spread'' of graph signals in both domains.
For example, \cite{agaskar_icassp} defines the spatial spread of any signal $\fv$ 
around a center vertex $i$ on a graph $\G$ as
\begin{equation}
 \Delta_{\G,i}^2(\fv) := \frac{1}{||\fv||_2^2}\sum_{j \in \Vc}[d_{\Gc}(i,j)]^2[f(j)]^2.
\label{eq:spatial_spread}
\end{equation}
Here, $\{[f(j)]^2/\norm{\fv}^2\}_{j=1,2,\ldots,N}$ can be interpreted 
as a probability mass function (pmf) of signal $\fv$,   
and  $\Delta_{\G,i}^2(\fv)$ 
is the variance of the geodesic distance function $d_{\Gc}(i,.):\Vc \to \RR$  at node $i$, in terms of this spatial pmf. The spatial spread of a graph signal can then be defined as 
\begin{align*}
\Delta^2_{\G}(\mathbf{f}):=\min_{i \in \V}\left\{\Delta^2_{\G,i}(\mathbf{f})\right\}.
\end{align*}

Similarly, 
the spectral spread of a graph signal can be defined as:
\begin{equation}
 \Delta_{\sigma}^2(\fv) := \min_{\mu {\in \Rbb_+}} \left\{\frac{1}{||\fv||_2^2}\sum_{\lambda \in \sigma(\Lcb)}\left[\sqrt{\lambda} - \sqrt{\mu}\right]^2\left[\hat f(\lambda)\right]^2\right\},
\label{eq:spectral_spread}
\end{equation}
where $\{[\hat f( \lambda)]^2/||\fv||_2^2\}_{\lambda=\lambda_0,\lambda_1,\ldots,\lambda_{\max}}$  is the pmf of $\fv$ 
across the spectrum of the Laplacian matrix, and $\sqrt{\mu}$ and $\Delta_{\sigma}^2(\fv)$ are the mean and 
variance 
of $\sqrt{\lambda}$, respectively, in the distribution given by 
this spectral pmf.\footnote{
Note that the definitions of spread 
presented here are heuristically defined and do not have a 
well-understood theoretical background. If the graph is not regular, the choice of which Laplacian matrix ($\Lcb$ or $\tilde \Lcb$) 
to use for computing spectral spreads also affects the results. The purpose of these definitions and the subsequent examples is to show that 
a trade-off exists between spatial and spectral localization in graph wavelets.}
If we do not minimize over all $\mu$ but rather fix $\mu=0$ and also use the normalized graph Laplacian matrix $\tilde \Lcb$ instead of $\L$,
the definition of spectral spread in (\ref{eq:spectral_spread}) 
reduces to the one proposed in \cite{agaskar_icassp}.

Depending on the application under consideration, other desirable features of a graph wavelet transform may include 
perfect reconstruction, 
critical sampling, orthogonal expansion,
and a multi-resolution decomposition \cite{narang_bipartite_prod}. 

In the remainder of this section, we categorize the existing graph transform designs and provide simple examples.  
The graph wavelet transform designs can broadly be divided 
into two types:
{\em vertex domain designs} and {\em graph spectral domain designs}. 
\subsection{Vertex Domain Designs} \label{Se:spatial}
The vertex domain 
designs of graph wavelet transforms are 
based on the spatial features of the graph, 
such as node connectivity and distances 
between vertices. Most of these localized transforms can be viewed as particular instances of filtering in the vertex domain, as in \eqref{Eq:vertex_filtering}, where
the output at each node can be computed 
from the samples within some 
$K$-hop neighborhood around the node. The graph spectral 
properties 
of these transforms are not  explicitly designed. Examples of vertex domain designs include
{\em random transforms}~\cite{wang}, 
{\em graph wavelets}~\cite{Crovella2003}, {\em lifting based wavelets} 
\cite{wagner,shen, narang_lifting_graphs}, and {\em tree wavelets}~\cite{gavish}.

The {\em random transforms}~\cite{wang} for unweighted graphs compute 
either a weighted 
average 
or  
a weighted difference at each node in the graph with respect to
a  
$k$-hop 
neighborhood around it. Thus, the filter at each node has 
a constant, non-zero weight $c$ within 
the $k$-hop neighborhood and zero weight outside, where the parameter $c$ 
is
chosen so as to guarantee invertibility of the transform. 

The {\em graph
wavelets} of Crovella and Kolaczyk \cite{Crovella2003} are functions $\psi_{k,i}: \Vc \to \mathbb{R}$, localized with
respect to a range of scale/location indices $(k, i)$, which 
at a minimum satisfy $\sum_{j \in \Vc}\psi_{k,i}(j) = 0$ 
(i.e. a zero DC response). This graph wavelet transform is described in more detail  
in Section \ref{Se:loc_example}.

Lifting based transforms for graphs~\cite{wagner,shen, narang_lifting_graphs}
are extensions of the lifting wavelets originally proposed  for 1D signals  by Sweldens~\cite{sweldens}. In this approach, the vertex set is 
first partitioned into sets of even and 
odd nodes,  
$\Vc = \V_{\Oc} \cup \V_{\Ec}$. 
Each odd node 
computes  
its {\em prediction} coefficient 
using its
own data and data from  
its even neighbors. Then each even node computes its
{\em update} coefficients 
using 
its own data and the  
prediction coefficients of its 
neighboring odd nodes. 

In~\cite{gavish}, Gavish et al. construct tree wavelets by building a balanced hierarchical tree from the data 
defined on graphs, and  
then generating orthonormal bases for the partitions defined 
at each level of the tree  using a modified version of the standard one-dimensional
wavelet filtering and decimation scheme.

\subsection{Graph Spectral Domain Designs}
The graph spectral domain designs of graph wavelets are
based on the spectral features of the graph, which are  
encoded, e.g., in  
the eigenvalues and eigenvectors of one of the graph matrices defined in Section \ref{Se:gsd}.
Notable examples in 
this category include 
diffusion wavelets~\cite{diffusion_wavelets,Maggioni_biorthogonal}, 
spectral graph wavelets~\cite{sgwt}, and graph quadrature mirror filter banks (graph-QMF filter banks)~\cite{narang_bipartite_prod}.
The general idea of the graph spectral designs is to construct bases that 
are localized in 
both the 
vertex and graph spectral domains. 

The diffusion wavelets~\cite{diffusion_wavelets,Maggioni_biorthogonal},
for example, are based  on compressed representations of powers of a diffusion operator, such as the one discussed in Example \ref{Ex:diffusion}. 
The localized basis functions at each resolution level are downsampled and then 
orthogonalized 
through a 
variation of the
Gram-Schmidt
orthogonalization 
scheme. 

The spectral graph wavelets of \cite{sgwt} are dilated, translated versions of a bandpass kernel designed in the graph spectral domain of 
the non-normalized graph Laplacian $\L$. They are discussed further in Section \ref{Se:loc_example}.

Another 
graph spectral design is the two-channel {\em graphQMF filter bank}
proposed for bipartite graphs
in
\cite{narang_bipartite_prod}. The resulting transform is orthogonal and critically-sampled, and also yields perfect reconstruction.
In this design, 
the analysis and synthesis filters at each scale
are 
designed using a single prototype transfer 
function $\hat h(\tilde \lambda)$, which satisfies:
\begin{equation}
 \hat h^2(\tilde \lambda) + \hat h^2( 2 - \tilde \lambda) = 2,
\end{equation}
 where $\tilde \lambda$ is an 
eigenvalue in the normalized graph Laplacian spectrum.
The design 
extends to any arbitrary graph via a {\em bipartite subgraph decomposition}. 

\subsection{Examples of Graph Wavelet Designs} \label{Se:loc_example}
In order to build more intuition about graph wavelets, we 
present some examples using one vertex domain design and one graph spectral domain design.

\begin{figure}[htb]
\centering
\includegraphics[width=2.5in]{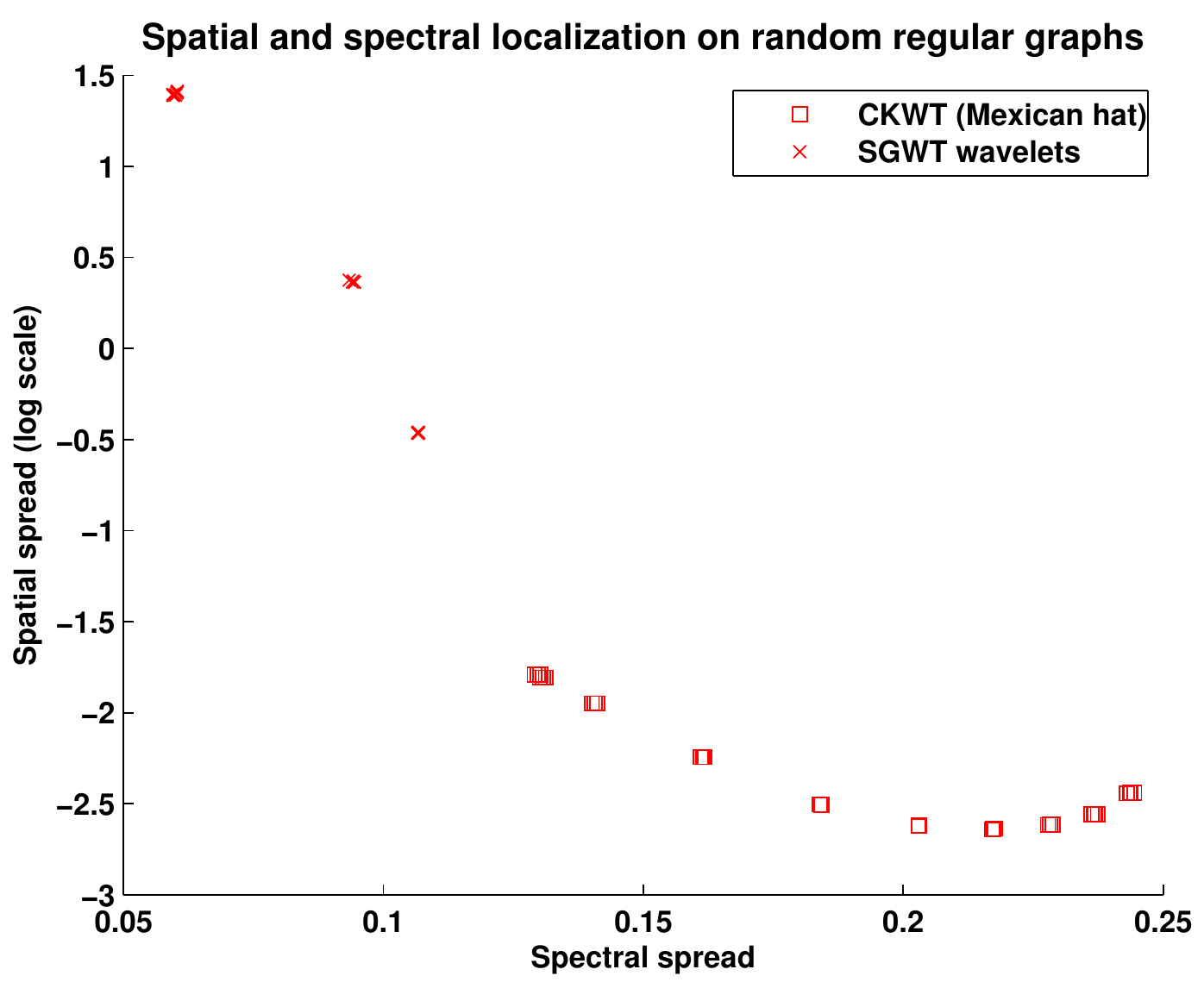}
\caption{\footnotesize {The average spatial and spectral spreads of two example wavelet transforms on $5$ instances of $d$-regular random graphs 
(size $N = 300$, degree $d = 5$).  The coordinates of each point in this figure are the average spatial and spectral spreads across all  wavelets at a given scale.
}}
\label{fig:spectral_spatial_spread}
\end{figure}

\begin{figure}[h]
\hfill
\begin{minipage}[b]{.30\linewidth}
   \centering
      \centerline{\small{$\mathbf{f}$}}  \vspace{.1cm}
   \centerline{\includegraphics[width=\linewidth]{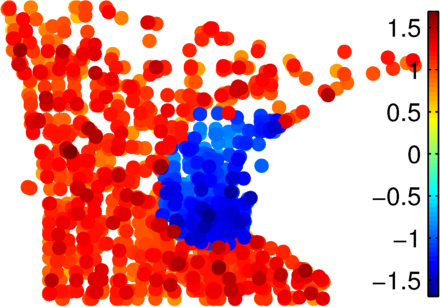}} 
   \centerline{\small{(a)}}
\end{minipage}
\hfill
\begin{minipage}[b]{.30\linewidth}
   \centering
         \centerline{\small{${\bf \Psi}^{CKWT}_{2}\fv$}}  \vspace{.1cm}
   \centerline{\includegraphics[width=\linewidth]{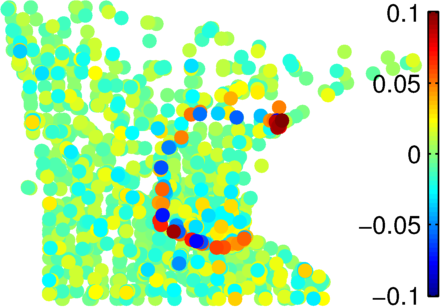}} 
      \centerline{\small{(b)}}
\end{minipage}
\hfill
\hfill
\begin{minipage}[b]{.30\linewidth}
   \centering
            \centerline{\small{${\bf \Psi}^{CKWT}_{4}\fv$}} \vspace{.1cm}
   \centerline{\includegraphics[width=\linewidth]{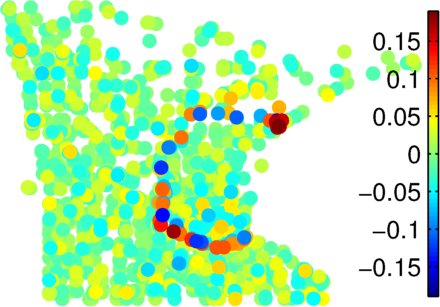}}
   \centerline{\small{(c)}}
\end{minipage}
\hfill
\hfill \vspace{.4cm} \\
\hfill
\begin{minipage}[b]{.35\linewidth}
   \centering
            \centerline{\small{${\bf \Psi}^{SGWT}_{scal}\fv$~}}  \vspace{.1cm}
   \centerline{\includegraphics[width=\linewidth]{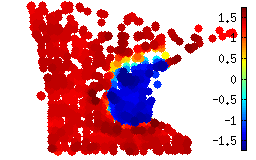}} 
   \centerline{\small{(d)}}
\end{minipage}
\hfill
\begin{minipage}[b]{.30\linewidth}
   \centering
            \centerline{\small{${\bf \Psi}^{SGWT}_{t_2}\fv$}}  \vspace{.1cm}
   \centerline{\includegraphics[width=\linewidth]{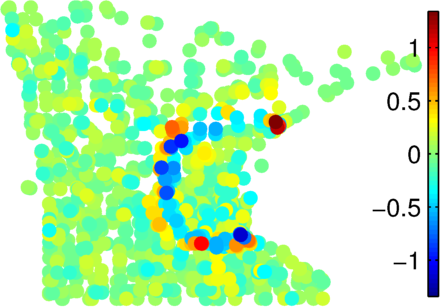}} 
      \centerline{\small{(e)}}
\end{minipage}
\hfill
\hfill
\begin{minipage}[b]{.30\linewidth}
   \centering
               \centerline{\small{${\bf \Psi}^{SGWT}_{t_4}\fv$}}  \vspace{.1cm}
   \centerline{\includegraphics[width=\linewidth]{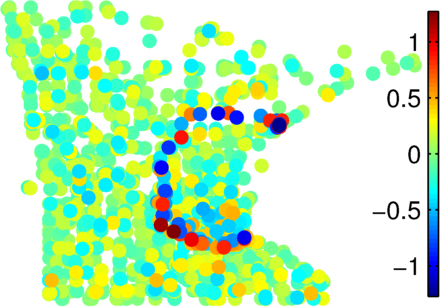}}
   \centerline{\small{~~~~(f)}}
\end{minipage}
\hfill
\hfill \\
\caption {(a) A piecewise smooth signal $\mathbf{f}$ with a severe discontinuity on the unweighted Minnesota graph. (b)-(c) Wavelet coefficients of two scales of the CKWT. 
(d) Scaling coefficients of the SGWT. (e)-(f) Wavelet coefficients of two scales of the SGWT. In both cases, the high-magnitude wavelet coefficients cluster around the discontinuity.}
  \label{Fig:wavelet}
\end{figure}

For the vertex domain design, we use the graph wavelet transform (CKWT) of Crovella and Kolaczyk \cite{Crovella2003} as an example.
These wavelets 
are 
based on the geodesic or shortest-path distance $d_{\Gc}(i,j)$. Define $\partial\N(i,\tau)$ to be the set 
of all vertices $j \in \Vc$ such that  $d_{\Gc}(i,j)= \tau$. Then 
 the wavelet function $\psi^{CKWT}_{k,i}: \Vc \to \mathbb{R}$ at scale $k$ and center vertex $i \in \Vc$ can be written as 
\begin{equation}
\psi^{CKWT}_{k,i}(j) = \frac{a_{k,\tau}}{|\partial\Nc(i,\tau)|},~\forall j \in \partial\N(i,\tau),
 \label{eq:crovella_filters_al}
\end{equation}
for some constants $\{a_{k,\tau}\}_{\tau=0,1,\ldots,k}$.
Thus, each wavelet is constant across all vertices $j \in \partial\N(i,\tau)$ that are the same distance from the center vertex $i$, and the value of the wavelet at the vertices in $\partial\N(i,\tau)$ depends on the distance $\tau$.
If $\tau > k$, $a_{k,\tau}=0$, so that for any $k$, the function $\psi^{CKWT}_{k,i}$ is exactly supported on 
a $k$-hop 
localized neighborhood around the center vertex $i$.
The constants $a_{k,\tau}$ in (\ref{eq:crovella_filters_al})
also satisfy $\sum_{\tau = 0}^{k} a_{k,\tau} = 0$, 
and can be computed from any continuous wavelet function $\psi^{[0,1)}(\cdot)$ supported on the interval $[0,1)$  
by taking  $a_{k,\tau}$ to 
be the average of $\psi^{[0,1)}(\cdot)$ on the sub-intervals $I_{k,\tau} = [\frac{\tau}{k+1}, \frac{\tau+1}{k+1}]$. 
In our examples in Figures \ref{fig:spectral_spatial_spread} and \ref{Fig:wavelet}, we take $\psi^{[0,1)}(\cdot)$ to be the {\em continuous Mexican hat} wavelet. We denote the entire graph 
wavelet transform at a given scale $k$ as ${\bf \Psi}_{k}^{CKWT} := [\psiv^{CKWT}_{k,1},\psiv^{CKWT}_{k,2},...\psiv^{CKWT}_{k,N}]$.

For the graph spectral domain design, we use the spectral graph wavelet transform (SGWT) of \cite{sgwt} as an example. The SGWT consists of one scaling function centered at each vertex, and $K$ wavelets centered at each vertex, at scales $\{t_1,t_2,\ldots,t_K\}\in \Rbb_+$. The scaling functions are translated low-pass kernels:
\begin{align*}
\psiv_{scal,i}^{SGWT}:=T_i \mathbf{h} = \hat{h}(\L)\boldsymbol{\delta}_i,
\end{align*}
where the generalized translation $T_i$ is defined in \eqref{Eq:new_translation}, and the kernel $\hat{h}(\lambda)$ is a low-pass filter.
The wavelet at scale $t_k$ and center vertex $i$ is defined as
\begin{align*}
\psiv_{t_k,i}^{SGWT}:=T_i \D_{t_k}\mathbf{g} = \widehat{\D_{t_k}{g}}(\L)\boldsymbol{\delta}_i,
\end{align*}
where the generalized dilation $\D_{t_k}$ is defined in \eqref{Eq:gen_dil}, and $\hat{g}(\lambda)$ is a band-pass kernel satisfying $\hat g(0) = 0$,  
$\lim_{\lambda \to\infty}\hat g(\lambda)=0$, and an admissibility condition \cite{sgwt}.
We denote the SGWT transform at scale $t_k$ as 
\begin{align*}
{\bf \Psi}_{t_k}^{SGWT} = [\psiv^{SGWT}_{t_k,1},\psiv^{SGWT}_{t_k,2},...\psiv^{SGWT}_{t_k,N}],
\end{align*} 
so that the entire transform ${\bf \Psi}^{SGWT}: \Rbb^N \rightarrow \Rbb^{N(K+1)}$ is given by
\begin{align*} 
{\bf \Psi}^{SGWT} = [{\bf \Psi}_{scal}^{SGWT};{\bf \Psi}_{t_1}^{SGWT};\ldots;{\bf \Psi}_{t_K}^{SGWT}].
\end{align*}

We now compute the spatial and spectral spreads of the two graph wavelet transforms 
presented above.
Unlike in the 
classical setting, the basis functions in a graph wavelet transform are not space-invariant; i.e., the spreads of two wavelets $\psiv_{k,i_1}$ and $\psiv_{k,i_2}$ at the same scale are not necessarily the same. 
Therefore, 
the spatial 
spread of a 
graph transform cannot be measured by computing the spreads of only one wavelet. 
In our analysis, we compute the spatial spread of a transform $\mathbf{\Psi}_k$ at a given scale $k$ by taking an average over all scale $k$ wavelet (or scaling) functions of the spatial spreads \eqref{eq:spatial_spread} around each respective center vertex $i$. 
Similarly, the spectral spread of the graph transform also changes with location. Therefore, we first compute 
\begin{equation}
 |\hat \Psi_{k}(\lambda)|^2 := \frac{1}{N}\sum_{i=1}^N |\hat \psi_{k,i}(\lambda)|^2,
\end{equation}
and then take $|\hat f(\lambda)|^2 = |\hat \Psi_{k}(\lambda)|^2$ in (\ref{eq:spectral_spread}) 
to compute the average spectral spread of ${\bf \Psi}_{k}$. 

The spatial and spectral spreads of both the CKWT and SGWT at different scales are shown 
in Figure~\ref{fig:spectral_spatial_spread}. 
The graphs used in this example are random $d$-regular graphs. 
Observe that in 
Figure~\ref{fig:spectral_spatial_spread},  
the CKWT wavelets are located to the right of the SGWT wavelets 
on the horizontal (spectral) axis, 
and below them on the vertical (spatial) axis, which implies 
that, in this example, 
the CKWT wavelets are less localized spectrally and more localized spatially  than the SGWT wavelets. 
This analysis provides an
empirical understanding of the trade-off between the spatial and spectral resolutions of 
signals defined on graphs. 

Next, to empirically demonstrate the ability of these graph wavelet transforms to efficiently represent piecewise smooth signals on graphs, we compute the graph wavelet coefficients of the piecewise smooth signal with a sharp discontinuity shown in Figure \ref{Fig:wavelet}(a) on the unweighted Minnesota road graph, where the color of a node represents the value of the signal at that vertex. We use the CKWT with scales $k=1,2,\ldots,10$, and the SGWT with 5 wavelet scales, as well as a scaling kernel. The bandpass wavelet kernel, scaling kernel, and values of the scales $t_1,t_2,t_3,$ and $t_4$ are all designed by the SGWT toolbox \cite{sgwt}. The CKWT wavelet coefficients as scales 2 and 4 are shown in Figures \ref{Fig:wavelet}(b) and \ref{Fig:wavelet}(c), and the SGWT scaling coefficients and wavelet coefficients at scales $t_2$ and $t_4$ are shown in Figures \ref{Fig:wavelet}(d)-(f), respectively.
Observe that for both transforms,  
the high-magnitude output coefficients are concentrated mostly near the discontinuity. This implies 
that these graph wavelet transforms are able to localize the {\em high-pass} information of the 
signal in the spatial domain, 
which the graph Fourier transform or other global transforms cannot do.

\section{Summary, Open Issues, and Extensions} \label{Se:discussion}

We presented a generic framework for processing data on graphs, and we surveyed recent developments in the area of graph signal processing.
 In particular, we 
 reviewed ways to generalize elementary operators such as filtering, convolution, and translation to graph setting. Such operations represent the core of graph signal processing algorithms, and they underly the localized, multiscale transforms we discussed in Section \ref{Se:transforms}. For many of the generalized operators defined in Section \ref{Se:ops} and the localized, multiscale transforms reviewed in Section \ref{Se:transforms}, classical signal processing intuition from Euclidean spaces can be fairly directly extended to the graph setting. For example, we saw in Section \ref{Se:frequency} how the notion of frequency extends nicely to the graph setting. However,
signals and transforms on graphs can also have surprising properties due to the irregularity of the data domain. Moreover, these are by no means the only conceivable ways to generalize these operators and transforms to the graph setting. Thus, quite a few challenges remain ahead. In this section, we briefly mention a few important open issues and possible extensions.
 
\subsection{Open Issues}
\begin{itemize}
\item Because all of the signal processing methods described in this paper incorporate the graph structure in some way, construction of the underlying graph is extremely important. Yet, relatively little is known about how the construction of the graph affects properties of the localized, multiscale transforms for signals on graphs. 
\item As mentioned in Section \ref{Se:other_graph}, it is not always clear when or why we should use
the normalized graph Laplacian eigenvectors, the non-normalized graph Laplacian eigenvectors, or some other basis as the graph spectral filtering basis.  
\item Similarly, in the vertex domain, a number of different distances, including the geodesic/shortest-path distance, the resistance distance \cite{klein}, the diffusion distance \cite{lafon_coarse}, and algebraic distances \cite{ron}, have useful properties, but it is not always clear which is the best to use in constructing or analyzing transform methods.
\item Transform operators are only useful in high-dimensional data analysis if the computational complexity of applying the operator and its adjoint scales gracefully with the size of the signal. This fact is confirmed, for example, by the prevalence of fast Fourier transforms and other efficient computational algorithms throughout the signal processing literature. Most of the transforms for signals on graphs involve computations requiring the eigenvectors of the graph Laplacian or the normalized graph Laplacian. However, it is not practical to explicitly compute these eigenvectors for extremely large graphs, as the computational complexity of doing so does not scale gracefully with the size of the graph. Thus, an important area of research is 
approximate computational techniques for signal processing on graphs. Efficient numerical implementations for certain classes of graph operators have been suggested using polynomial approximations \cite{smola,sgwt,shuman_DCOSS_2011} and Krylov methods \cite{hancock}, but plenty of numerical issues remain open, including, e.g., a fast graph Fourier transform implementation.
\item In Euclidean data domains, there is a deep mathematical theory of approximation linking properties of classes of signals to properties of their wavelet transform coefficients (see, e.g., \cite{donoho_theory}). A major open issue in the field of signal processing on graphs is how to link  structural properties of graph signals and their underlying graphs to properties (such as sparsity and localization) of the generalized operators and transform coefficients. Such a theory could inform transform designs, and help identify which transforms may be better suited to which applications. One issue at the heart of the matter is the need to better understand localization of signals in both the vertex and graph spectral domains. As discussed briefly in Section \ref{Se:transforms}, even defining appropriate notions of spreads in these domains is highly non-trivial. 
Moreover, unlike in the classical Euclidean settings, the graph Laplacian eigenvectors are often highly non-localized, making it more difficult to precisely identify the trade-off between resolution in the vertex domain and resolution in the graph spectral domain. 
Agaskar and Lu \cite{agaskar_icassp} have begun to define such localization notions and study the resolution trade-off. 
\end{itemize} 

\subsection{Extensions}
The signal processing techniques we have described are focused on extracting information from a static signal on a static, weighted, undirected graph. Some clear extensions of this framework include: 1) considering directed graphs, as is done for example in \cite{chung_directed}; 2) considering time series of data on each vertex in a graph;  
3) considering a time-varying series of underlying graphs, as is done for example in \cite{Lee_SAMPTA}; or any combination of these.

Finally, while the number of new analytic techniques for signals on graphs has been steadily increasing over the past decade, the application of these techniques to real science and engineering problems is still in its infancy. We believe the number of potential applications is vast, and hope to witness increased utilization of these important theoretical developments over the coming decade.

\balance 

\bibliographystyle{IEEEtran}
\bibliography{bib_whitepaper}

\end{document}